\documentclass[12pt]{article}
\usepackage{hyperref,slashed,graphicx,color,amsmath,amssymb}
\usepackage[numbers,sort&compress]{natbib}

\usepackage{subfig}

\begin{document}

\vskip 1cm
\begin{center}
{ \bf\large
Majorana neutrinos with point interactions
\vskip 0.2cm}
\vskip 0.3cm
{Chengfeng Cai and Hong-Hao Zhang\footnote{Email: zhh98@mail.sysu.edu.cn}}
\vskip 0.3cm
{\it \small School of Physics and Engineering, Sun Yat-Sen University, Guangzhou
510275, China}
\end{center}
\vspace{0.3cm}
\begin{center}
{\bf Abstract}
\end{center}

We propose a realistic model with Majorana neutrinos in the framework of unifying the three generations of fermions by point interactions in an extra dimension.
This model can simultaneously explain the origin of fermion generations, fermion masses and mixing, and the smallness of the masses of Majorana neutrinos.
We show that there are two mechanisms working together to suppress the neutrino masses significantly, so we do not have to introduce a very large extra-dimension cut-off scale. One is the type-I seesaw mechanism and the other is the overlap integration of localized lepton wave functions. A singlet scalar with an exponential-like VEV plays a central role in these two mechanisms. For consistency in this model we introduce a $U(1)'$ gauge symmetry, which will be broken by the singlet scalar. Parameters of our model can fit the masses and flavor mixing data well. These parameters can also predict all CP violating phases including the Majorana ones and accidentally rescue the proton from decay.

\newpage
\tableofcontents

\section{Introduction}
The recent discovery of the Higgs boson is a great success for the Standard Model (SM) of particle physics. In the SM, the masses of weak gauge bosons and fermions are generated by the Higgs mechanism, which predicts the existence of a CP-even scalar particle, and finally this only scalar boson was discovered at the Large Hadron Collider (LHC) in 2012 \cite{Aad:2012tfa,Chatrchyan:2012ufa}.

However, many people believe that the SM should not be the finale of particle physics. One of the reasons is that it cannot explain the large hierarchy of fermion masses. In the SM, all fermion masses, mixing angles and CP phases are free parameters. If one looks at the mass spectrum of fermions, one will find a significant hierarchy between different generations. The hierarchy between quark sector and lepton sector is even worse.

In the original version of the SM, the neutrino masses are assumed to be zero. However, to explain the oscillation phenomena observed in experiments, the neutrinos have to be massive. Similar to the way used in the SM to give fermions masses, it can make neutrinos massive by introducing right-handed neutrinos which couple to the Higgs field through Yukawa terms. But this way is quite unnatural due to the large hierarchy. A cosmological observation from Planck set a $0.23$ eV upper bound for the sum of the three generations of neutrinos \cite{Ade:2013zuv}. It leads to about 11 order of magnitude hierarchy between the Yukawa coupling of top quark and the neutrinos. This unnaturalness indicates us a strong motivation to go beyond the SM.

There are three types of seesaw mechanisms to explain the smallness of neutrino masses. The type-I seesaw introduces right-handed neutrinos coupled with the left-handed leptons through Yukawa interactions, and then the Majorana masses of the left-handed neutrinos will be generated by a higher dimensional operator and be suppressed by the heavy Majorana masses of the right-handed ones \cite{Minkowski:1977sc,Yanagida:1979as,GellMann:1980vs}. The type-II seesaw introduces triplet scalars coupled with the left-handed lepton doublets, and the vacuum expectation value (VEV) of the scalar will be suppressed by its large quadratic masses \cite{Mohapatra:1979ia,Schechter:1980gr,Schechter:1981cv}. The type-III seesaw is similar to the type-I, but it introduces heavy triplet leptons \cite{Foot:1988aq}.
All these mechanisms usually need a high seesaw energy scale, for example the Grand Unification Theory (GUT) scale, to suppress the induced Majorana masses of the left-handed neutrinos.

Besides the seesaw mechanisms, an alternative way to explain the masses hierarchy naturally is to enlarge the spacetime dimension. One interesting case is the thick wall model \cite{ArkaniHamed:1999dc}, in which fermions have Gaussian wave functions of the 5th dimension coordinate and their locations are determined by their 5-dimensional (5D) masses. When two fermion wave functions are separated slightly, their overlap integration with the Higgs VEV profile will be suppressed exponentially, then a large hierarchy structure between fermions can be naturally obtained.
Another fascinating case is the Randall-Sundrum model \cite{Randall:1999ee,Grossman:1999ra}, in which right-handed neutrinos localize near a hidden brane, while the other fermions and the Higgs field are confined on a visible brane. Thus the right-handed neutrinos interact with the other fields weakly, and they only have tiny masses.

Recently, a new extra-dimension model \cite{Fujimoto:2012wv,Fujimoto:2014fka} was proposed to unify the 3 fermion generations. The model introduces 5D fermion fields living in an extra-dimensional interval or circle with several point interactions (\textit{i.e.} 0-thickness branes). For each 5D fermion, there are three independent modes between branes. They behave as three generations, and the hierarchy between generations is achieved by coupling the 5D fermion field to a scalar field which has an exponentially increasing extra-dimensional coordinate-dependent VEV. This specific VEV can be generated by imposing Robin's boundary conditions on the scalar at two boundaries of the 5th dimension (see more details on the phase structures in \cite{Fujimoto:2011kf}). In addition, a twisted boundary condition is imposed on the Higgs doublet to create CP violating phases for both quark and lepton sectors \cite{Fujimoto:2013ki}. In Ref.~\cite{Fujimoto:2014fka}, a 5D singlet neutrino field (which has a right-handed chiral neutrino 0-mode) is introduced to construct Dirac masses terms for neutrinos, and the smallness of neutrino masses are obtained from a proper arrangement of the point interaction positions. A stringent constraint on the model with a set of fitted parameters is to suppress the proton decay rates. By a rough analysis with some baryon number violating dimension-8 operators, the cut-off $\Lambda\sim L^{-1}$ is estimated to be as large as the GUT scale ($10^{15}$ GeV).

In this paper, we discuss a possibility to extend the model of Ref.~\cite{Fujimoto:2014fka} to a Majorana neutrino case and to avoid the large cut-off scale. To implement this, we need a Majorana mass term of the singlet neutrinos. A naive trial is to write down an explicit Majorana mass term for the singlet neutrino fields and their charge conjugation. However, it fails since the equations of motion for the singlet neutrinos no longer respects the so called quantum mechanical supersymmetry (QMSUSY) which is important for acquiring chiral zero modes \cite{Witten:1981nf,ArkaniHamed:1999dc}.
The existence of a Majorana mass term implies that the lepton number is no longer a conserved quantity, and thus a dimension-7 effective operator $\overline{L}\sigma^2H^\ast H^\dag\sigma^2L^c$ may appear in the Lagrangian in principle. Here $L(x,y)$ is the 5D lepton doublet field, $H(x,y)$ is the 5D Higgs doublet, and the power counting is achieved in 5D spacetime. But this effective operator can induce large Majorana masses of the left-handed neutrinos after the electroweak symmetry breaking. To avoid large neutrino masses which violates the experimental bounds, it requires either a high cut-off scale or a very small coupling constant for this term.

To overcome this problem and to forbid the harmful explicit Majorana masses terms at the same time, we introduce a new $U(1)'$ gauge symmetry. If we let the singlet neutrino field $N_R$ and the combination $H^\dag\sigma^2L^c$ be $U(1)'$ charged, \textit{none} of these two annoying terms, $\overline{N_R^c}N_R$ and $\overline{L}\sigma^2H^\ast H^\dag\sigma^2L^c$, can survive, since they are doubly $U(1)'$ charged.
In other words, we increase the symmetry of the model to prohibit the unwanted operators like $\overline{L}\sigma^2H^\ast H^\dag\sigma^2L^c$.

We need a natural way to realize the experimentally acceptable Majorana masses for neutrinos.
Consider another dimension-7 operator $\left(\Phi^{\ast}\right)^2\overline{N_R^c}N_R+\mathrm{h.c.}$, which is available
if the $U(1)'$ charge of the singlet scalar $\Phi$ is assigned to be the same as that of the singlet neutrino $N_R$. Obviously, This term can contribute to a Majorana mass for the right-handed neutrino 0-mode when the scalar $\Phi$ obtains a non-zero VEV and break the $U(1)'$ gauge symmetry. The 5D scalar $\Phi(x,y)$ is initially introduced to realize the hierarchy of the three generations of quarks and leptons, and it is imposed on the Robin's boundary condition to get a VEV, $\langle \Phi(y)\rangle$, as an exponential-like function of the extra dimensional coordinate $y$. This VEV has the effect of killing two birds with one stone. If the 0-thickness branes' positions of singlet neutrino are chosen appropriately, that is, if the third generation singlet neutrino wave function has a big overlap with the large value side of $\langle \Phi(y)\rangle$, it can obtain a mass which is much larger than the Dirac masses. This large mass turns on the type-I seesaw mechanism to lower the neutrino masses further.

To make our model self-consistent, we set the $U(1)'$ charge of each field agreeing with the anomaly free conditions \cite{Rizzo:2006nw}. We also consider the constraint from the proton decay. By some simple analysis with the dimension-8 baryon number violating operators, we see that for our best-fit parameters the proton will not decay.
So it is not necessary to let the cut-off energy be the GUT scale in this model.

An outline of the paper is as follows. In section \ref{sect2}, we will summarize some key elements of the model and building a realistic model in the framework. We also discuss what are the problems of introducing an explicit Majorana mass term. In section \ref{sect3}, we discuss how to generalize the model to include Majorana neutrinos and how the seesaw mechanism works with a few TeV extra-dimension energy scale. We will also fit the data of leptons and do some discussion. Section \ref{sect4} is a summary. In Appendix \ref{app1},\ref{app2},\ref{app3}, we supply some mathematical details of the discussion in section \ref{sect2}.
\section{The model\label{sect2}}

To begin with, let us summarize some general setups of this model (with some mathematical details reviewed in Appendix \ref{app1}) \cite{Fujimoto:2012wv,Fujimoto:2014fka}:
\begin{itemize}
\item The spacetime is extended by a finite size of space-like extra-dimension, \textit{i.e.} an interval or a circle. The mode expansion is made as usual and the lowest modes, \textit{i.e.}, the zero modes, are regarded as the SM particles. The mass gap between the 1st. K.K. modes and the zero modes is roughly the inverse of the 5th dimensional size. In many extra dimension models, the mass scale is at least around the energy scale of the LHC experiment.

\item In the free field limit, there is a quantum mechanical supersymmetry (QMSUSY) between the left-handed and right-handed components of 5D fermion \cite{ArkaniHamed:1999dc,Witten:1981nf,Cooper:1994eh}. This symmetry ensures that the left-handed and right-handed modes at the same level have equal masses. Thus, their 4D parts can be separated from the 5th-dimension-coordinate-dependent parts, and can form a Dirac fermion satisfying the 4D Dirac equation. In particular, for the zero mode, the symmetry together with the Dirichlet boundary conditions implies that one of the chiral spinors should vanish and the other one is massless. This is the method of generating chiral zero modes.

\item An important ingredient for unifying  generations is the point interaction \cite{Fujimoto:2012wv,Fujimoto:2014fka}, which can be regarded as a Delta-function-like interaction. This specific interaction is located at a point in the 5th dimension and results in the Dirichlet boundary condition for the 5D fermion. If we introduce two interacting points, then they will separate the interval at extra dimension into three pieces. The modes living in different pieces are independent from each other although they come from the same 5D fermion field. These different modes can be regarded as different generations.

 \item To achieve the hierarchy among generations, a singlet scalar field $\Phi(x,y)$ is introduced to couple with 5D fermions.
A Robin's boundary condition on the 5D scalar will force its VEV $\langle\Phi(y)\rangle$ to be $y$-dependent as
\begin{eqnarray}\label{solution}
\langle\Phi(y)\rangle=\frac{\nu}{\mathrm{cn}(\sqrt{\frac{\lambda}{2}}\frac{\mu}{k}(y-y_0),k)}
\end{eqnarray}
where the function $\mathrm{cn}(x,k)$ is the Jacobi elliptic function of $x$ with index $k$,
and $k,\mu,\nu$ are defined as
\begin{eqnarray}
\left\{\begin{array}{l}k^2=\frac{\mu^2}{\mu^2+\nu^2}\\
\mu^2=\frac{M^2}{\lambda}(1+\sqrt{1+\frac{4\lambda|Q|}{M^4}})\\
\nu^2=\frac{M^2}{\lambda}(\sqrt{1+\frac{4\lambda|Q|}{M^4}}-1)\end{array}\right.
\end{eqnarray}
with $Q,y_0$ being constants of integration determined by $L_{\pm}$.
A study of this singlet scalar with Robin's boundary condition can be found in Ref.~\cite{Fujimoto:2011kf}. An important result in their study is that $\Phi(x,y)$ can couple with gauge fields corresponding to some group, such as a $U(1)'$ group. This symmetry will break if $L<L_c=\frac{1}{|M|}\tanh^{-1}\left(\frac{|M|(L_++L_-)}{1+M^2L_+L_-}\right)$ \cite{Fujimoto:2011kf,Fujimoto:2012wv}. Usually we use the condition $M^2<\frac{1}{L_{\mathrm{max}}},L_{\mathrm{max}}=\mathrm{max}(L_+,L_-)$, which is sufficient but not necessary.
\end{itemize}

When we proceed to construct a realistic model comparable with experiments, some special settings are also needed \cite{Fujimoto:2012wv,Fujimoto:2014fka,Fujimoto:2013ki}. The requirements are briefly listed as follows:
\begin{enumerate}
\item The 5th dimension need to be a circle ($S^1$). This is a part of the requirements from the flavour mixing behavior of the SM. And it is also consistent with the twisted boundary condition setting of the Higgs doublet.

\item We need to specify the 5D matter fields with appropriate boundary conditions. In the quark sector, we should introduce an electroweak $SU(2)$ doublet quark $Q(x,y)=(U_L(x,y)\ D_L(x,y))^T$, and two singlets quarks $U_R(x,y)$ and $D_R(x,y)$.
For the doublet $Q$, we use a Dirichlet boundary condition $P_RQ=0$ at $y=L_0^{(q)}=0,L_1^{(q)},L_2^{(q)}$ so that its zero modes are left-handed, while for the singlets $U_R$ and $D_R$, we use Dirichlet boundary conditions $P_LU_R=0$ at $y=L_0^{(u)},L_1^{(u)},L_2^{(u)}$ and $P_LD_R=0$ at $y=L_0^{(d)},L_1^{(d)},L_2^{(d)}$ so that their zero modes are right-handed. Note that in general $L_i^{(q)}$ are different from $L_i^{(u)}$ and  $L_i^{(d)}$. This is necessary for flavor mixing structure. For the lepton sector, the situation is similar to the quark case. We just replace the quark doublet by a lepton doublet and the up and down type quark singlet by neutrino and charged lepton singlet.
\item We need a Higgs doublet $H(x,y)$ to couple with fermion fields through Yukawa couplings.
Of course it should acquire non-zero VEV $\langle H\rangle$ to break the electroweak symmetry.
A special treatment is to impose a twisted boundary condition on $H(x,y)$ as $H(y+L)=e^{i\theta}H(y)$ \cite{Fujimoto:2013ki}.
This twisted boundary condition will make the VEV $\langle H\rangle$ get $y$ dependent phase as $\langle H(y)\rangle=\frac{v}{\sqrt{2L}}e^{\frac{i\theta}{L}y}$, then its overlap integration with fermions' wave functions will produce CP phases for CKM or PMNS matrices.
\end{enumerate}
As an example, the detailed treatment of the quark sector are presented in Appendix \ref{app2}. We also fit the parameters of quark sector independently and list them in Table.~\ref{quarkbf}. The fitting will fix the $M$ parameter from the singlet scalar $\Phi$ and the $\theta$ from the Higgs $H$, and they will be regarded as input data for the lepton case.\par
Before going to the next section to discuss our treatment of the lepton sector. It will be helpful to ask what's wrong if we just write down an explicit Majorana mass term? We will discuss this briefly as follows, and supply more details in Appendix \ref{app3}.\par
One problem of this naive trial is that Majorana mass term will modify the equation of the motion for the 5D fermion. This modification breaks the QMSUSY between the left-handed and right-handed components in the E.O.M. As we mentioned previously, generating chiral zero modes rely on this symmetry.\par
Another problem with this naive trial is that since we are going to break the lepton number conservation explicitly, then in principle we should also include an operator as $\overline{L}\sigma^2H^\ast H^\dag\sigma^2L^c$ which has the same dimension with the terms we used to generate the Dirac masses for leptons. After the Higgs acquires a non-zero VEV, this operator will generate Majorana masses for the left-handed neutrino zero modes. Then a fine-tuning is needed when we diagonalize the neutrino mass matrix to obtain sub-eV masses.

\section{The lepton sector}\label{sect3}
\subsection{$U(1)'$ symmetry and type-I seesaw}
For the lepton sector, we introduce an $SU(2)$ doublet $L=(N_L(x,y),E_L(x,y))^T$, and singlets $N_R(x,y),E_R(x,y)$.
When we consider the structure of our model, the lepton number is not necessary to be preserved.
The most famous model which violate lepton number is the type-I seesaw \cite{Mohapatra:1979ia}.
In type-I seesaw a Majorana mass term for the right-handed neutrino is introduced.
If the Majorana mass $M_R$ is extremely large comparing to the Dirac mass $m_D^{(\nu)}$, then after diagonalize the mass matrix, a mass for the three lightest neutrinos taking the form $-m_D^{(\nu)}M_R^{-1}m_D^{(\nu)T}$ will be suppressed significantly. But as we discussed in the section \ref{sect2}, an explicit Majorana mass term is not allowed to exist.
We will assign a $U(1)'$ charge to $N_R$ to forbid such a troublesome term to keep the chiral 0-mode, and then use the VEV of the scalar $\Phi$ to create the Majorana masses for the right-handed neutrino 0-mode.\par
As we have mentioned in section \ref{sect2}, the $\overline{L}\sigma^2H^\ast H^\dag\sigma^2L^c$ operator will bring us a problem of fine-tuning. To solve this problem it will be forbidden by the $U(1)'$ symmetry if we let $\overline{L}i\sigma^2H^\ast$ to be charged.
All these indicate that we would better add the $U(1)'$ symmetry into the model.
Then to justify the model, we should put some constraints to the undetermined $U(1)'$ charges.\par
The gauge group in our model is now $SU(3)_C\times SU(2)_L\times U(1)_Y\times U(1)'$.
Let's denote the representation of all left-handed zero modes in the form $(N_{c,i},N_{w,i},Y_i,Q'_i)$, where $N_{c,i}$ and $N_{w,i}$ denote the dimensions of $SU(3)$ and $SU(2)$ representation(conjugated representation with a bar) of $i$-th field, while $Y_i$ and $Q'_i$ denote the $U(1)$ hyper-charge and $U(1)'$ charge of $i$-th field. $N_{c,i},N_{w,i},Y_i$ for each type of field are just the same as in the Standard Model. $Q'_i$s for each type of field are unknown variables and will be determined later. We list the representations for fermions in table \ref{reps}
\begin{table}[!htbp]
\caption{Gauge group representations for fermions}\label{reps}
\centering
\begin{tabular}{|c|c|c|c|}
\hline
Fields    &$q$  &$u_R^c$  & $d_R^c$  \\
\hline
Reps.  &$(3,2,1/6,Q'_q)$    &$(\bar{3},1,-2/3,Q'_u)$  &$(\bar{3},1,1/3,Q'_d)$  \\
\hline \hline
Fields    & $l$ & $\nu_R^c$ & $e_R^c$ \\
\hline
Reps.  &$(1,2,-1/2,Q'_l)$ & $(1,1,0,Q'_n)$ & $(1,1,1,Q'_e)$ \\
\hline
\end{tabular}
\end{table}
Now the covariant derivatives for each field are
\begin{eqnarray}
\begin{array}{l}
  D^{(Q)}_N=\partial_N-ig_sG^i_Nt^i-igW^a_NT^a-i\frac{1}{6}g'B_N-iQ'_qg_cC_N\\
  D^{(U)}_N=\partial_N-ig_sG^i_Nt^i-i\frac{2}{3}g'B_N+iQ'_ug_cC_N\\
  D^{(D)}_N=\partial_N-ig_sG^i_Nt^i+i\frac{1}{3}g'B_N+iQ'_Dg_cC_N\\
  D^{(L)}_N=\partial_N-igW^a_NT^a+i\frac{1}{2}g'B_N-iQ'_lg_cC_N\\
  D^{(N)}_N=\partial_N+iQ'_ng_cC_N\\
  D^{(E)}_N=\partial_N+ig'B_N+iQ'_eg_cC_N\\
  D^{(H)}_N=\partial_N-igW^a_NT^a-i\frac{1}{2}g'B_N-iQ'_hg_cC_N\\
  D^{(\Phi)}_N=\partial_N-iQ'_\phi g_cC_N
\end{array}
\end{eqnarray}
Where the $C_N$ is the gauge field corresponding to $U(1)'$ and $g_c$ is the gauge coupling.
There are 6 constraints of $Q'_i$ come from the consideration of anomaly free \cite{Rizzo:2006nw}.
They are
    \begin{eqnarray}
    \begin{array}{r}
      2Q'_q+Q'_u+Q'_d=0\\
      3Q'_q+Q'_l=0\\
      6(\frac{1}{6})^2Q'_q+3[-(\frac{2}{3})^2Q'_u+(\frac{1}{3})^2Q'_d]+2(-\frac{1}{2})^2Q'_l+Q'_e=0\\
      6Q^{'3}_q+3[Q^{'3}_u+Q^{'3}_d]+2Q^{'3}_l+Q^{'3}_e+Q^{'3}_n=0\\
      6\cdot\frac{1}{6}Q^{'2}_q+3[-\frac{2}{3}Q^{'2}_u+\frac{1}{3}Q^{'2}_d]+2(-\frac{1}{2})Q^{'2}_l+Q^{'2}_e=0\\
      6Q'_q+3[Q'_u+Q'_d]+2Q'_l+Q'_e+Q'_n=0\end{array}
    \end{eqnarray}
It seems that we have 6 equations for 6 variable, but actually only 4 of them are independent.
We rewrite $Q'_i$s in terms of $Q'_l$ and $Q'_e$ as follows
\begin{eqnarray}\label{qi}
  \left\{\begin{array}{l}Q'_q=-\frac{1}{3}Q'_l\\Q'_u=-\frac{2}{3}Q'_l-Q'_e\\Q'_d=\frac{4}{3}Q'_l+Q'_e\\Q'_n=-2Q'_l-Q'_e\end{array}\right.
\end{eqnarray}
Then when we choose a set $(Q'_l,Q'_e)$, all the other variables are determined.
For our purpose, we will impose more theoretical constraints on $Q'_i$s.
One is that we need Yukawa terms as
\begin{eqnarray}
\begin{array}{ll}
  \Phi\overline{Q}(i\sigma^2H^\ast)U_R,&\Phi^\ast\overline{Q}HD_R,\\
  \Phi\overline{L}(i\sigma^2H^\ast)N_R,&\Phi^\ast\overline{L}HE_R.\end{array}
\end{eqnarray}
to be gauge invariant. Assign a $U(1)'$ charge $Q'_h$ to $H$ and $Q'_\phi$ to $\Phi$, and use \eqref{qi} finally we find the only constraint is
\begin{eqnarray}\label{yuk}
  Q'_l+Q'_e-Q'_h+Q'_\phi=0
\end{eqnarray}
Another important constraint is to let $Q'_\phi=Q'_n$ so that $\Phi^2\overline{N_R^c}N_R$ is gauge invariant, or let $Q'_\phi=-Q'_n$ so that $\Phi^{\ast2}\overline{N_R^c}N_R$ is gauge invariant.
Then we replace $Q'_\phi$ by $\pm Q'_n$ in \eqref{yuk} and use \eqref{qi}, we obtain $Q'_h=-Q'_l$ for $\Phi^2\overline{N_R^c}N_R$ or $Q'_h=3Q'_l+2Q'_E$ for $\Phi^{\ast2}\overline{N_R^c}N_R$.
Remember that we want $\overline{L}(i\sigma^2H^\ast)$ to be $U(1)'$ charged and it requires that $Q'_h\neq -Q'_l$, so only $Q'_h=3Q'_l+2Q'_e$ corresponding to $\Phi^{\ast2}\overline{N_R^c}N_R$ is allowed.
Of course, we should have $Q'_n\neq0$ to kill the explicit Majorana mass term for singlet neutrino and this requires that $Q'_e\neq-2Q'_l$.
The other constraints may come from experimental considerations but that is beyond the scope of this article.\par
There are still many possible choices of $Q'_i$s and we only list three interesting candidates which are similar to \cite{Appelquist:2002mw,Coriano:2014mpa}:
\begin{enumerate}
\item $U_{R}$:~~$Q'_l=Q'_q=0,Q'_u=1,Q'_d=-1,Q'_e=-1,Q'_n=1,Q'_h=-2,Q'_\phi=-1$.
\item $U_{B-L}$:~~$Q'_q=\frac{1}{3},Q'_u=Q'_d=-\frac{1}{3},Q'_l=-1,Q'_n=Q'_e=1,Q'_h=-1,Q'_\phi=-1$.
\item $U_{\chi}$:~~$Q'_q=\frac{1}{5},Q'_u=\frac{1}{5},Q'_d=-\frac{3}{5},Q'_l=-\frac{3}{5},Q'_n=1,Q'_e=\frac{1}{5},Q'_h=-\frac{7}{5},Q'_\phi=-1$.
\end{enumerate}
The mass term of zero-mode leptons will be generated by
\begin{eqnarray}\label{yukawaact}
  \mathcal{L}_{yuk}&=&-\int dy[\mathcal{Y}^{(n)}\Phi(y)\overline{L}(i\sigma^2H^\ast)N_R+\mathcal{Y}^{(e)}\Phi^\ast(y)\overline{L}HE_R+h.c.]\nonumber\\
  &&-\frac{1}{2}\int dy[y^{(m)}\Phi^{\ast2}\overline{N_R^c}N_R+h.c.]
\end{eqnarray}
where $\mathcal{Y}^{(n)},\mathcal{Y}^{(e)}$ and $y^{(m)}$ are couplings with dimension $-2$.
After the $U(1)'$ and $SU(2)\times U(1)$ breaking, two terms in the first line generate Dirac mass matrices for charged leptons and neutrinos and the term in the second line generate a Majorana mass matrix for right-handed neutrinos.\par
\begin{figure}[!htbp]
\begin{center}
     \includegraphics[width=0.5\textwidth]{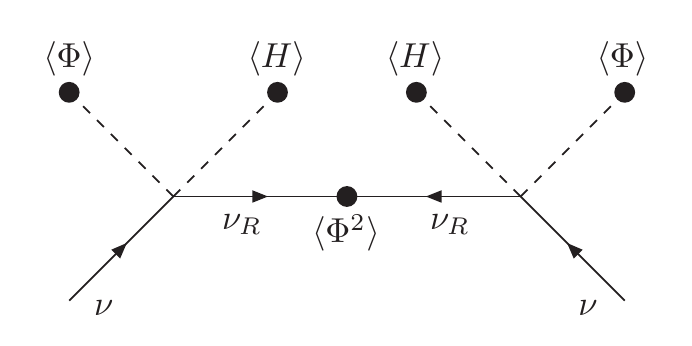}
\end{center}\caption{A diagrammatic description of eq.~\eqref{Mnu}}\label{feyndiag}
\end{figure}

Imposing Dirichlet boundary conditions on fermion fields, twisted boundary condition on Higgs doublet and Robin boundary condition on $\Phi$, we can expand fields in modes and finally obtain their profiles:
\begin{eqnarray}
L&=&\sum_{i=1}^3\left(\begin{matrix}f_{l_{iL}^{(0)}}(y)\nu_{iL}^{(0)}(x)\\
f_{l_{iL}^{(0)}}(y)e_{iL}^{(0)}(x)\end{matrix}\right)+\,\mathrm{(KK~modes)},\nonumber\\
E_R&=&\sum_{i=1}^3f_{e_{iR}^{(3)}}(y)e_{iR}^{(0)}(x)+\,\mathrm{(KK~modes)},\nonumber\\
N_R&=&\sum_{i=1}^3f_{\nu_{iR}^{(3)}}(y)\nu_{iR}^{(0)}(x)+\,\mathrm{(KK~modes)},\qquad N_R^c=C\overline{N_R}^T,\\
f_{l_{iL}^{(0)}}(y)&=&N_{iL}^{(l)}e^{M_L(y-L_{i-1}^{(l)})}\theta(y-L_{i-1}^{(l)})\theta(L_{i}^{(l)}-y),\nonumber\\
f_{e_{iR}^{(0)}}(y)&=&N_{iR}^{(e)}e^{-M_E(y-L_{i-1}^{(e)})}\theta(y-L_{i-1}^{(e)})\theta(L_{i}^{(e)}-y),\nonumber\\
f_{\nu_{iR}^{(0)}}(y)&=&N_{iR}^{(\nu)}e^{-M_N(y-L_{i-1}^{(\nu)})}\theta(y-L_{i-1}^{(\nu)})\theta(L_{i}^{(\nu)}-y)\nonumber
\end{eqnarray}
where $N_{iL}^{(l)},N_{iR}^{(e)},N_{iR}^{(\nu)}$ are normalization constants.
Substituting these profiles into \eqref{yukawaact}, we get the Dirac mass matrices and Majorana mass matrix:
\begin{eqnarray}
\begin{array}{l}
m_{ij}^{(e)}=\int dy\mathcal{Y}^{(e)}\frac{v}{\sqrt{2L}}\langle\Phi(y)\rangle f_{l_{iL}^{(0)}}(y)f_{e_{jR}^{(0)}}(y)e^{\frac{i\theta y}{L}},\\
m_{D,ij}^{(n)}=\int dy\mathcal{Y}^{(n)}\frac{v}{\sqrt{2L}}\langle\Phi(y)\rangle f_{l_{iL}^{(0)}}(y)f_{\nu_{jR}^{(0)}}(y)e^{-\frac{i\theta y}{L}},\\
M_{R,ij}=y^{(m)}\int_{L_{i-1}^{(\nu)}}^{L_i^{(\nu)}}dy\langle\Phi(y)\rangle^2f_{\nu_{iR}^{(0)}}(y)f_{\nu_{jR}^{(0)}}(y)\end{array}
\end{eqnarray}
Obviously, $M_R$ is a diagonal matrix since the integration only involves the profile of $N_R$.
Now we write the chiral zero modes in Weyl basis:
\begin{eqnarray}
\nu_{iL}^{(0)}\to\nu_{iL,a},\qquad e_{iL}^{(0)}\to e_{iL,a},\qquad \nu_{iR}^{(0)}\to\nu_{iR}^{\dag,\dot{a}}
\end{eqnarray}
where $a,\dot{a}$ are indices of Weyl spinors. Then for neutrinos we can represent the mass term as
\begin{eqnarray}\label{neutrinomass}
\mathcal{L}_{mass}^{(\nu)}=-\frac{1}{2}(\nu_{iL,\dot{a}}^\dag\ \nu_{iR,\dot{a}}^\dag)\left(\begin{matrix}0&m_{D,ij}^{(n)}\\(m_{D,ij}^{(n)})^T&M_{R,ij}\end{matrix}\right)\left(\begin{matrix}\nu_{jL}^{\dag,\dot{a}}\\ \nu_{jR}^{\dag,\dot{a}}\end{matrix}\right)+h.c.
\end{eqnarray}
Following Xing's parametrization and discussion \cite{Xing:2011ur}, we introduce a $6\times6$ unitary matrix $\mathcal{U}$ to transform the mass eigenstates to flavor states. $\mathcal{U}$ can be decomposed into
\begin{eqnarray}
  \mathcal{U}=\left(\begin{matrix}\textbf{1}&\textbf{0}\\ \textbf{0}&U_0\end{matrix}\right) \left(\begin{matrix}A&R\\S&B\end{matrix}\right) \left(\begin{matrix}V_0&\textbf{0}\\ \textbf{0}&\textbf{1}\end{matrix}\right)=\left(\begin{matrix}AV_0&R\\U_0SV_0&U_0B\end{matrix}\right)
\end{eqnarray}
where $V_0$ and $U_0$ are $3\times3$ unitary matrices and $A,B,R,S$ are $3\time3$ matrices under the unitary conditions:
\begin{eqnarray}
\begin{array}{l}
AA^\dag+RR^\dag=BB^\dag+SS^\dag=\textbf{1},\\
AS^\dag+RB^\dag=AR^\dag+S^\dag B=\textbf{0},\\
A^\dag A+S^\dag S=B^\dag B+R^\dag R=\textbf{1}\end{array}
\end{eqnarray}
We can use $\mathcal{U}$ to diagonalize the mass matrix in \eqref{neutrinomass}:
\begin{eqnarray}\label{diag}
  \mathcal{U}^\dag\left(\begin{matrix}0&m_{D,ij}^{(n)}\\(m_{D,ij}^{(n)})^T&M_{R,ij}\end{matrix}\right)\mathcal{U}^\ast=\left(\begin{matrix}\widehat{M}_\nu&\textbf{0}\\ \textbf{0}&\widehat{M}_N\end{matrix}\right)
\end{eqnarray}
where $\widehat{M}_\nu$ and $\widehat{M}_N$ are diagonal matrices: $\widehat{M}_\nu=Diag\{m_1,m_2,m_3\}$ are very small while $\widehat{M}_N=Diag\{M_1,M_2,M_3\}$ should be very large.
Finally we can find approximately
\begin{eqnarray}\label{Mnu}
\widehat{M}_\nu\simeq-V_0^\dag(m_{D,ij}^{(n)}M_R^{-1}(m_{D,ij}^{(n)})^T)V_0^\ast
\end{eqnarray}
The minus sign can be absorbed into charged lepton basis.
Remember that at the beginning of this section, we use the $U(1)'$ symmetry to kill the $\overline{L}\sigma^2H^\ast H^\dag\sigma^2L^c$ dimension-7 operator. When the $U(1)'$ symmetry breaks spontanuously, this term comes back by connecting two Yukawa interaction with an internal Majorana sterile neutrino line.
A diagrammatic description of eq.~\eqref{Mnu} is shown in Fig.~\ref{feyndiag}.
Thus, the smallness of this Majorana mass is natural.\par
Masses $m_D^{(n)}$ and $M_R$ are determined by model parameters and then we can use Takagi diagonalization with the unitary matrix $V_0$ to diagonalize the symmetric complex matrix $m_{D,ij}^{(n)}M_R^{-1}(m_{D,ij}^{(n)})^T$ \cite{Dreiner:2008tw}.
The PMNS matrix $V_0$ can be parametrized as
\begin{eqnarray}\label{V0}
V_0=\left(\begin{matrix}c_{12}c_{13}&\hat{s}_{12}^\ast c_{13}&\hat{s}_{13}^\ast\\-\hat{s}_{12}c_{23}-c_{12}\hat{s}_{13}\hat{s}_{23}^\ast&c_{12}c_{23}-\hat{s}_{12}^\ast\hat{s}_{13}\hat{23}^\ast&c_{13}\hat{s}_{23}^\ast\\ \hat{s}_{12}\hat{s}_{23}-c_{12}\hat{s}_{13}c_{23}&-c_{12}\hat{s}_{23}-\hat{s}_{12}^\ast\hat{s}_{13}c_{23}&c_{13}c_{23}\end{matrix}\right)
\end{eqnarray}
where $c_{ij}\equiv\cos\theta_{ij},\hat{s}_{ij}\equiv e^{i\delta_{ij}}\sin\theta_{ij}$, $\theta_{ij}$s are mixing angles of active neutrino and $\delta_{ij}$s are CP phase angles(3 for Majorana neutrinos).\par
\begin{figure}[!htbp]
\begin{center}
     \includegraphics[width=0.8\textwidth]{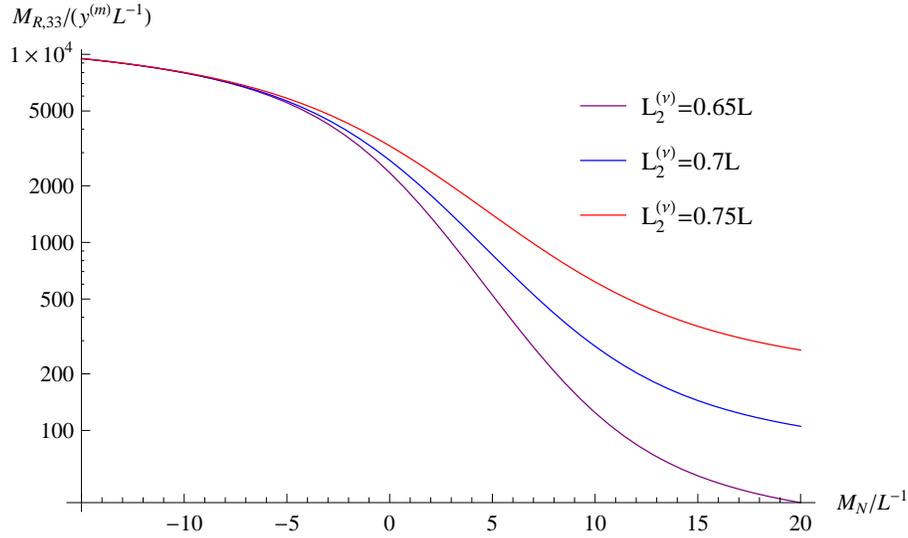}
\end{center}\caption{$M_{R,33}$ vs. $M_N$ with $L_3^{(\nu)}\to1$ and $L_{2}^{(\nu)}=0.65,0.7,0.75$}\label{MR33}
\end{figure}

As we know, to suppress the neutrino masses to sub-eV with the seesaw mechanism, we need extremely large $M_R$s.
Interestingly, this can be achieved by the exponentially increasing behavior of the VEV $\langle\Phi(y)\rangle$.
The matrix element $M_{R,ij}$ can be estimated as follows
\begin{eqnarray}
M_{R,ij}&=&y^{(m)}\int_{L_{i-1}}^{L_i}dy\langle\Phi(y)\rangle^2f_{\nu_{iR}^{(0)}}(y)f_{\nu_{jR}^{(0)}}(y)\nonumber\\
&\approx&y^{(m)}N_{iR}^{(\nu)2}\delta_{ij}\int_{L_{i-1}^{(\nu)}}^{L_i^{(\nu)}}dy\nu^2\cosh^2(M(y-y_0))e^{-2M_N(y-L_{i-1}^{(\nu)})}\nonumber\\
&\approx&\delta_{ij}\frac{y^{(m)}|Q|}{M_N^2-M^2}\left[\frac{M_N^2}{M^2}-1
+\frac{M_N^2}{M^2}\cosh(2M(L_{i-1}^{(\nu)}-y_0))\right.\nonumber\\
&&+\frac{M_N}{M}\sinh[2M(L_{i-1}^{(\nu)}-y_0)]
+e^{-2M_N(L_i^{(\nu)}-L_{i-1}^{(\nu)})}\left(\frac{M_N^2}{M^2}-1\right.\nonumber\\
&&+\left.\left.\frac{M_N^2}{M^2}\cosh[2M(L_{i}^{(\nu)}-y_0)]
+\frac{M_N}{M}\sinh[2M(L_{i}^{(\nu)}-y_0)]\right)\right]
\end{eqnarray}
We plot the 3rd element of the diagonal, $M_{R,33}$, as a function in terms of $M_N$ and let $L_3^{(\nu)}\to L,L_{i-1}^{(\nu)}=0.65L,0.7L,0.75L$ in Fig.~\ref{MR33}.
This function increases when $L_{2}^{(\nu)}$ increases or $\tilde{M_N}$ decreases, and we find that if $\tilde{M_N}<15,L_{2}^{(\nu)}\sim0.75$ then $M_{R,33}$ can be as large as $500L^{-1}\sim10000L^{-1}$.

Apparently there are hierarchies $M_{R,11}\ll M_{R,22}\ll M_{R,33}$ and one may worry that some element of matrix $m_D^{(\nu)}M_R^{-1}(m_D^{(\nu)})^T$ is not suppressed by $M_{R,33}$, but by $M_{R,11}$ instead.
So we show the explicit expression of $m_D^{(\nu)}M_R^{-1}(m_D^{(\nu)})^T$ as follows
\begin{eqnarray}
\left(\begin{matrix}\frac{m_{11}^2}{M_{R,11}}+\frac{m_{12}^2}{M_{R,22}}+\frac{m_{13}^2}{M_{R,33}} & \frac{m_{11}m_{21}}{M_{R,11}}+\frac{m_{12}m_{22}}{M_{R,22}}+\frac{m_{13}m_{23}}{M_{R,33}} & \frac{m_{11}m_{31}}{M_{R,11}}+\frac{m_{12}m_{32}}{M_{R,22}}+\frac{m_{13}m_{33}}{M_{R,33}} &\\
\frac{m_{11}m_{21}}{M_{R,11}}+\frac{m_{12}m_{22}}{M_{R,22}}+\frac{m_{13}m_{23}}{M_{R,33}} & \frac{m_{21}^2}{M_{R,11}}+\frac{m_{22}^2}{M_{R,22}}+\frac{m_{23}^2}{M_{R,33}} & \frac{m_{11}m_{31}}{M_{R,11}}+\frac{m_{12}m_{32}}{M_{R,22}}+\frac{m_{13}m_{33}}{M_{R,33}} &\\
\frac{m_{11}m_{31}}{M_{R,11}}+\frac{m_{12}m_{32}}{M_{R,22}}+\frac{m_{13}m_{33}}{M_{R,33}} & \frac{m_{11}m_{31}}{M_{R,11}}+\frac{m_{12}m_{32}}{M_{R,22}}+\frac{m_{13}m_{33}}{M_{R,33}} & \frac{m_{31}^2}{M_{R,11}}+\frac{m_{32}^2}{M_{R,22}}+\frac{m_{33}^2}{M_{R,33}}
\end{matrix}\right)
\end{eqnarray}
Then we see that all terms contain $m_{33}$ (which assumed to be the largest element of Dirac mass matrix) are suppressed by $M_{R,33}$.
Also note that $m_{11},m_{22}$, etc. are usually much smaller than $m_{33}$, so their suppression don't need masses as large as $M_{R,33}$.

In conclusion, thanks to the exponential-like VEV of the scalar, although our scale $L^{-1}$ is only about order of TeV, it is still possible to lower the neutrino mass $m_D^{(\nu)}M_R^{-1}(m_D^{(\nu)})^T$ to sub-eV with the Majorana mass $M_R$.

\subsection{Numerical results and discussion}
Since we have fitted the parameters of the scalar $\Phi$ and $H$ in the quark case (see Appendix\ref{app2}), we set them fixed in the lepton fitting. Although we extend the gauge group in this model, but it will not affect the parameters we obtained in the quark case. Note that the parameter $y^{(m)}$ comes into the fitting only in a combination $\frac{\mathcal{Y}^{(n)}}{\sqrt{y^{(m)}}}$, so we will not treat $y^{(m)}$ and $\mathcal{Y}^{(n)}$ separately. In our fitting, we only consider the normal hierarchy of neutrino mass.\par
The recent experiment data of leptons have been used in our fitting are listed in the following,
\begin{itemize}
\item Masses of charged leptons: $m_e=(0.510998928\pm1.1\times10^{-8})\textrm{MeV},\ m_\mu=(105.6583715\pm3.5\times10^{-6})\textrm{MeV},\ m_\tau=(1776.82\pm0.16)\textrm{MeV}$ \cite{Beringer:1900zz}.
\item Mass squared difference between two generations: $\Delta m_{31}^2=(2.473\pm0.069)\times10^{-3}\textrm{eV}^2,\Delta m_{21}^2=(7.5\pm0.19)\times10^{-5}\textrm{eV}^2$
\cite{GonzalezGarcia:2012sz}.
\item Mixing angles: $\sin^2\theta_{12}=0.302\pm0.012,\sin^2\theta_{23}=0.413\pm0.032,\sin^2\theta_{13}=0.0227\pm0.0024$
\cite{GonzalezGarcia:2012sz}.
\end{itemize}
Since there are more free parameters than data, we only show one set of the possible parameters. They are listed in Table~\ref{leptonbf}.
\begin{table}[!htbp]
\caption{Best fit parameters for leptons}\label{leptonbf}
\centering
\begin{tabular}{|c|c|c|c|}
\hline
$L^{(l)}_0$    &$L^{(l)}_1$  &$L^{(l)}_2$  & $M_L$\\
\hline
$0.378389L$ &$0.670380L$   &$0.908743L$   &$-11.792317L^{-1}$\\
\hline
\hline
$L^{(n)}_0$    &$L^{(n)}_1$  &$L^{(n)}_2$  & $M_N$\\
\hline
$0.062289L$     &$0.515437L$   &$0.741436L$   &$13.293167L^{-1}$\\
\hline
\hline
$L^{(e)}_0$    &$L^{(e)}_1$  &$L^{(e)}_2$  & $M_E$\\
\hline
$0.317799L$     &$0.448665L$   &$0.701578L$   &$36.580911L^{-1}$\\
\hline
\hline
$\frac{\tilde{\mathcal{Y}}^{(e)}v}{\sqrt{2}}$    &$\frac{\tilde{\mathcal{Y}}^{(n)}v}{\sqrt{2\tilde{y}^{(m)}}}\sqrt{\frac{\textrm{TeV}}{L^{-1}}}$    &-&-\\
\hline
$0.317575$GeV     &$0.000319953$GeV     &-&-\\
\hline
\end{tabular}
\end{table}
If we assume that $\tilde{y}^{(m)}\sim O(1)$ (a parameter with a tilde means it is scaled by $L$ to be dimensionless), then we can see that the hierarchy between $\mathcal{Y}^{(e)}$ and $\mathcal{Y}^{(n)}$ is about 3 order of magnitude which is acceptable. Notice that when $L^{-1}$ has larger magnitude such as $10$TeV or $100$TeV, $\frac{\tilde{\mathcal{Y}}^{(n)}v}{\sqrt{2}}$ may get closer to $\frac{\tilde{\mathcal{Y}}^{(e)}v}{\sqrt{2}}$.
If we compare the Yukawa couplings with that for the quark sector in Table.~\ref{quarkbf},
we will find that $\mathcal{Y}^{(e)}$ has the same order with $\mathcal{Y}^{(d)}$.
So no hierarchy of the Yukawa couplings between quarks and leptons.
All lepton 5D masses $M_L$,$M_E$ and $M_N$ are O(10) up to the scale $L^{-1}$ which are also seemed natural.\par
This set of parameters will give
\begin{itemize}
\item Masses of charged leptons:\\
$m_e=0.510999$ MeV, $\ m_\mu=105.65837$ MeV, $\ m_\tau=1776.79963$ MeV.\\
They all deviate the experimental value less than 0.01\% as the fitting required.
\item Masses of neutrinos: \\
$m_1=0.005074$ eV, $\ m_2=0.010092$eV, $\ m_3=0.049868$eV.\\
Comparing with the data, the mass squared differences between the 1st and 3rd generation deviates the experimental one about 0.5\%, while the mass squared differences between the 1st and 2nd generation deviates the experimental one about 1.5\%.
\item Masses of sterile neutrinos:\\
$M_1=1.2144\textrm{GeV}\frac{\tilde{y^{(m)}}L^{-1}}{\textrm{TeV}},\ M_2=4.9870\textrm{TeV}\times\frac{\tilde{y^{(m)}}L^{-1}}{\textrm{TeV}},\ M_3=358.8498\textrm{TeV}\frac{\tilde{y^{(m)}}L^{-1}}{\textrm{TeV}}$.\\
Both $\tilde{y^{(m)}}$ and the scale $L^{-1}$ are undetermined. But we can see that if $\tilde{y^{(m)}}L^{-1}\sim O(1\sim10\textrm{TeV})$, the lightest sterile neutrino can be produced by the LHC, and since it interact weakly with other particles, it may only contribute to a little part of the missing $E_t$.

\item Mixing angles:\\
$\sin^2\theta_{12}=0.30315,\sin^2\theta_{23}=0.4359,\sin^2\theta_{13}=0.0221$.\\
They all deviate the experimental value less than 6\%.
\item CP phases:\\
$\delta_{12}=0.1944,\delta_{23}=1.2796,\delta_{13}=3.0716$.
\end{itemize}
We can also calculate the effective Majorana mass as:
\begin{eqnarray}
\langle m_{\beta\beta}\rangle\equiv|\sum_km_kU_{ek}^2|=7.43\textrm{meV}
\end{eqnarray}
This quantity is related to the double-beta decay which now have limit $\langle m_{\beta\beta}\rangle\lesssim120\sim250$meV (90\%CL.) \cite{Gando:2012zm}. Not surprisingly that our result is far from the experimental limit since the masses of active neutrinos are all smaller than $100$ meV.\par
We can also estimate the mass of gauge field $C_\mu$ as follows:
\begin{eqnarray}
\frac{1}{2}m_c^2C^\mu C_\mu&=&\int dy\frac{1}{L}g_c^{2}\langle\Phi\rangle^2C^\mu C_\mu
\end{eqnarray}
which implies
\begin{eqnarray}
m_c^2&=&2\tilde{g}_c^{2}\int dy\langle\Phi\rangle^2
=2\tilde{g}_c^{2}\frac{2|Q|}{M}\left(\frac{L}{2}+\frac{L}{4\tilde{M}}(\sinh(2\tilde{M}+2\tilde{M}\tilde{y}_0)
-\sinh2\tilde{M}\tilde{y}_0)\right)\nonumber\\
&\approx&\frac{\tilde{g}_c^{2}|\tilde{Q}|L^{-2}}{2\tilde{M}^2}e^{2\tilde{M}(1-\tilde{y}_0)}
\end{eqnarray}
which further leads to
\begin{eqnarray}
m_c\approx\tilde{g}_c\frac{\sqrt{|\tilde{Q}|}L^{-1}}{\sqrt{2}\tilde{M}}e^{\tilde{M}(1-\tilde{y}_0)}
\approx(124\cdot\tilde{g}_c)\textrm{TeV}\left(\frac{L^{-1}}{\textrm{TeV}}\right)
\end{eqnarray}
So for $\tilde{g}_c\approx0.1\sim1,L^{-1}\approx1\sim100$ TeV we have $m_c\approx10\sim10000$ TeV.
Notice that there is another mixing effect if $H$ is $U(1)'$ charged.
When EW symmetry breaks, there will be a mass term involving $Z$ and $C$ \cite{Rizzo:2006nw}, then to obtain the mass eigenvalues we shall diagonalize a mass matrix in $(Z,C)$ basis as
\begin{eqnarray}
\mathcal{M}^2=\left(\begin{matrix}m_Z^2&\beta m_Z^2\\ \beta m_Z^2&m_c^2\end{matrix}\right)
\end{eqnarray}
where $\beta$ is a factor about $O(1)$ or less. Since our $m_c^2$ is apparently much larger than $m_Z^2$, so the mixing would not be significant and the $\rho=\frac{m_Z^2}{m_1^2}$ is very closed to 1, where $m_1$ is the smaller mass eigenvalue.
Notice that this heavy gauge field will also significantly suppressed the effective coupling of some process mediated by it.
The effective coupling which is similar to the Fermi constant $G_c\sim\frac{\tilde{g}_c^{2}}{m_c^2}=\frac{1}{(124L^{-1})^2}\approx\frac{G_F}{(500L^{-1}/\textrm{TeV})^2}$ is much smaller than $G_F$, so this process will not change the whole amplitude.\par
\begin{figure}[!htbp]
\begin{center}
     \includegraphics[width=0.45\textwidth]{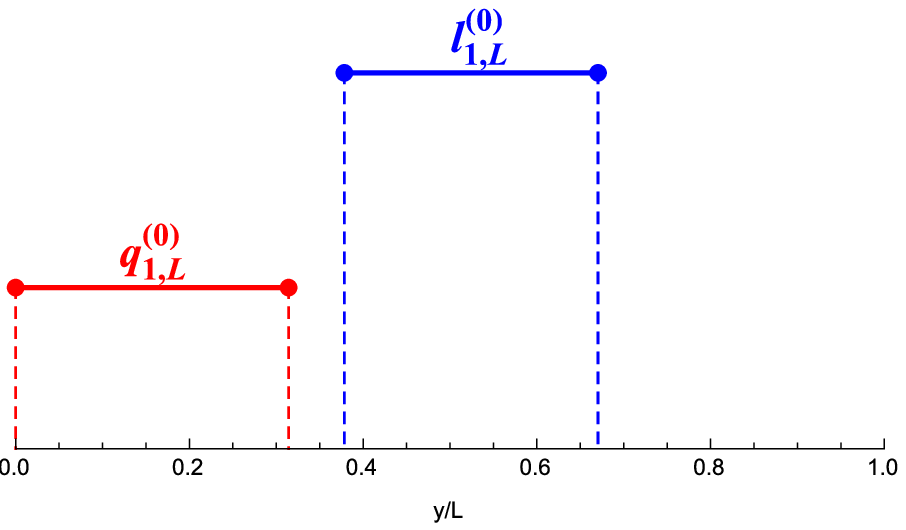}
     \includegraphics[width=0.45\textwidth]{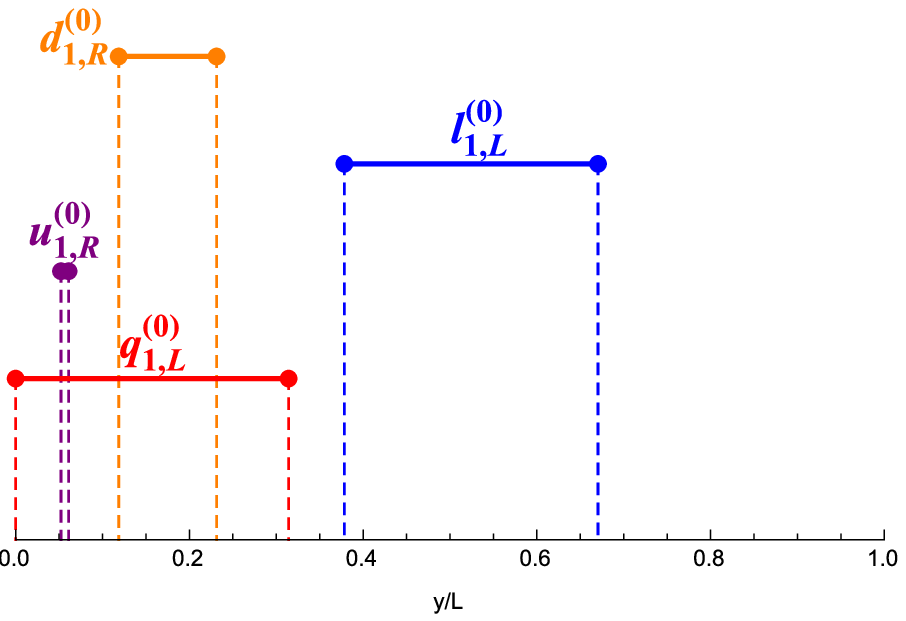}\\
     \includegraphics[width=0.45\textwidth]{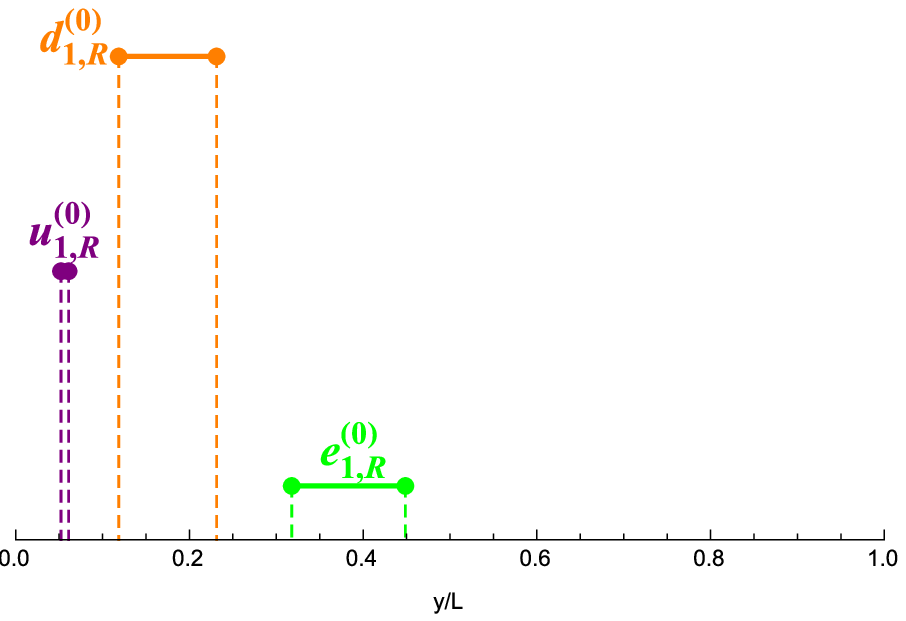}
     \includegraphics[width=0.45\textwidth]{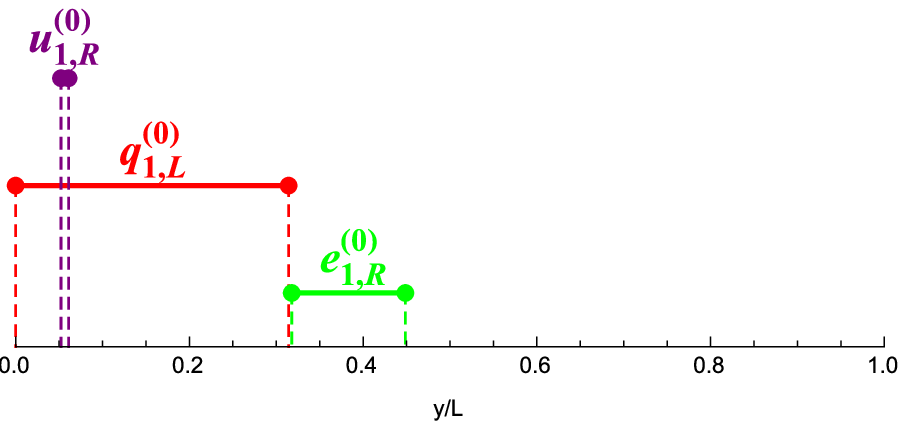}
\end{center}\caption{The domains of the first generation wave functions. The left-top is for the operator $QQQL$; the right-top is for the $DUQL$; the left-bottom is for the $UDEU$; while the right-bottom is for the $QQUE$.}\label{prdecay}
\end{figure}

Interestingly, given the parameters shown in Tables.~\ref{leptonbf}, \ref{quarkbf}, we do not need to worry about the constraints from the proton decay. Following the analysis of \cite{Fujimoto:2014fka}, the dimension-eight operators lead to proton decay are $QQQL,DUQL,UDEU$ and $QQUE$.
We show the domains of the first generation wave functions which involved in the operators in Fig.~\ref{prdecay} .
We find that for each operator, there are at least two domains do not overlap, and thus the integration vanishes.

\section{Summary}\label{sect4}
In this paper, we have discussed the possibility to generalize the model constructed in Ref.~\cite{Fujimoto:2012wv,Fujimoto:2014fka} to a Majorana neutrino case. The extra dimension scale $L^{-1}$ is about several TeV, which seems far from the scale for seesaw mechanism and is unlikely to explain the small neutrino masses naturally. But we note that the smallness of neutrino masses can be a synthesized effect of the Type-I seesaw and the overlap integration of the localized lepton wave functions. We find that a 5D scalar $\Phi$ with an exponentially warped VEV, which was initially introduced in Refs.~\cite{Fujimoto:2012wv,Fujimoto:2014fka} to generate a hierarchy between generations, can also be used to generate large Majorana masses for the neutrino right-handed 0-modes.
The strategy is to let $\Phi$ couple with singlet neutrino field in the manner $\Phi^{\ast2}\overline{N_R^c}N_R$.
When $\Phi$ acquires a non-zero vacuum expectation value, $\langle\Phi(y)\rangle^2$, which exponentially depends on the extra dimension coordinate $y$, will be extremely large near $y=L$ so that the third generation of right-handed neutrino will be very heavy and turn on the seesaw mechanism. At the same time, if the positions of the 0-thickness branes and the 5D bulk mass $M_N$ are properly chosen, the overlap integration of the left handed and right handed neutrino wave functions will be also smaller than that of the charged leptons. Both of these effects work together, and they can significantly suppress the neutrino masses.\par
To justify the model, it is necessary to add a $U(1)'$ gauge symmetry into the model. This symmetry prohibits some troublesome terms like $\overline{L}\sigma^2H^\ast H^\dag\sigma^2L^c$ and the explicit Majorana terms.
When $\Phi$ obtains a non-zero vacuum expectation value, the $U(1)'$ symmetry will break spontaneously.
Since the mass of the $U(1)'$ gauge boson is very large, it will not change the prediction significantly.
For consistency, we also discuss how the anomaly cancellation conditions constrain the $U(1)'$ charge of each
field. The numerical results of our model parameters have no significant hierarchy among them.
They can fit all masses and flavor mixing data very well.
We use this set of parameters to calculate some observable quantities such as the effective Majorana mass, and we find it is consistent with the double-beta decay experiments. Our parameters also rescue us from the stringent proton-decay constraint on the cut-off scale.

\vspace{0.6cm}

\section*{Acknowledgments}

This work is supported in part by the National Natural Science Foundation
of China (NSFC) under Grant Nos. 11375277, 11410301005 and 11005163,  the Fundamental
Research Funds for the Central Universities, and Sun Yat-Sen
University Science Foundation.

\appendix
\section{The general setup of the framework}\label{app1}
In this appendix we briefly review the extra dimension model with point interactions.
The basic setup is to let all fields live in 5D spacetime and have point interactions with some 0-thickness branes \cite{Fujimoto:2012wv,Fujimoto:2014fka}.
The point interaction means a $\delta$-function-potential-like interaction which vanishes everywhere except at a point in the 5th dimension \cite{Cheon:2000tq,Nagasawa:2008an,Fujimoto:2012wv}.

The action of a 5D fermion field $\Psi(x,y)$ is given by \cite{Fujimoto:2012wv}
\begin{eqnarray}\label{act1}
S=\int d^4x\int dy\bar{\Psi}(x,y)(i\Gamma^M\partial_M+M_F)\Psi(x,y)
\end{eqnarray}
where $M_F$ is the 5D bulk mass, and the $\Gamma$ matrices obey the Clifford algebra $\{\Gamma_M,\Gamma_N\}=-2\eta_{MN}$ with the 5D metric $\eta_{MN}=\mathrm{diag}\{-1,1,1,1,1\}$ and the indices $M,N=0,1,2,3,5$ and $\mu,\nu=0,1,2,3$. An explicit representation of the $\Gamma$ matrices is $\Gamma^\mu=\gamma^\mu$ and $\Gamma^y=-i\gamma_5=\gamma^0\gamma^1\gamma^2\gamma^3$.
The variation of the action \eqref{act1} is:
\begin{eqnarray}\label{variation-1}
\delta S&=&\int d^4x\int dy\left[\delta\bar{\Psi}(i\Gamma^M\partial_M+M_F)\Psi
+\bar{\Psi}(i\Gamma^M\partial_M+M_F)\delta\Psi\right]\nonumber\\
&=&\int d^4x\int dy\left[\delta\bar{\Psi}(i\Gamma^M\partial_M+M_F)\Psi
-\bar{\Psi}(i\Gamma^M\overleftarrow{\partial}_M-M_F)\delta\Psi
+\partial_M(\bar{\Psi}i\Gamma^M\delta\Psi)\right]\nonumber\\
\end{eqnarray}
Thus, $\delta S/\delta\bar{\Psi}=0$ implies the equation of motion (EOM) for $\Psi$:
\begin{eqnarray}\label{eom}
(i\Gamma^M\partial_M+M_F)\Psi=
\left(\begin{matrix}-\partial_y+M_F&i\sigma^\mu\partial_\mu\\
i\bar{\sigma}^\mu\partial_\mu&\partial_y+M_F\end{matrix}\right)
\left(\begin{matrix}\Psi_L\\\Psi_R\end{matrix}\right)=0
\end{eqnarray}
where the field $\Psi(x,y)$ has been decomposed into the left-handed and right-handed components $\Psi_{L,R}=P_{L,R}\Psi=[(1\mp\gamma_5)/2]\Psi$ in the chiral representation of Dirac matrices $\gamma^\mu$. Taking complex conjugate of eq.~\eqref{eom} gives the EOM for $\bar{\Psi}$: $\bar{\Psi}(i\Gamma^M\overleftarrow{\partial}_M-M_F)=0$. Substituting it and \eqref{eom} into \eqref{variation-1} and taking $\delta S=0$, we obtain
\begin{eqnarray}
0=\int d^4x\int dy\partial_M(\bar{\Psi}\Gamma^M\delta\Psi)
=\int d^4x\int dy[\partial_\mu(\bar{\Psi}\Gamma^\mu\delta\Psi)+\partial_y(\bar{\Psi}\Gamma^y\delta\Psi)]
\end{eqnarray}
Since the integral of the 4D total divergence vanishes: $\int d^4x\partial_\mu(\bar{\Psi}\Gamma^\mu\delta\Psi)=0$, we have
\begin{eqnarray}
\int dy\partial_y(\bar{\Psi}\Gamma^y\delta\Psi)=0 \label{boundary-1}
\end{eqnarray}
which, as we have seen, is required for the consistency of the EOMs for $\Psi$ and $\bar{\Psi}$.

Now let us consider a toy model, in which the extra 1-dimensional space is an interval with length $L$ and in the 5th dimension there are 3 boundary points assigned as $0,~L_1(<L),~L$, respectively. In this case, eq.~\eqref{boundary-1} implies
\begin{eqnarray}
0&=&\int_0^L dy\partial_y(\bar{\Psi}\Gamma^y\delta\Psi)
=\left(\int_0^{L_1-\epsilon}+\int_{L_1+\epsilon}^L\right) dy\partial_y(\bar{\Psi}\Gamma^y\delta\Psi)\nonumber\\
&=&(\bar{\Psi}\Gamma^y\delta\Psi)\bigg|_{y=L}-(\bar{\Psi}\Gamma^y\delta\Psi)\bigg|_{y=0}
+(\bar{\Psi}\Gamma^y\delta\Psi)\bigg|_{y=L_1-\epsilon}
-(\bar{\Psi}\Gamma^y\delta\Psi)\bigg|_{y=L_1+\epsilon}\label{boundary-2}
\end{eqnarray}
where $\epsilon$ is a positive infinitesimal length. A sufficient condition to satisfy eq.~\eqref{boundary-2} is to let the term vanish at all the boundary points:
\begin{eqnarray}
\bar{\Psi}\Gamma^y\delta\Psi=i(\Psi_R^\dag\delta\Psi_L-\Psi_L^\dag\delta\Psi_R)
=0\qquad (\text{at~~} y=0,~L_1\pm\epsilon, ~L)\label{boundary-3}
\end{eqnarray}
It is sufficient to satisfy eq.~\eqref{boundary-3} by imposing the Dirichlet boundary condition
\begin{equation}\label{Dirichlet}
\Psi_R=0\quad\text{or}\quad\Psi_L=0\qquad (\text{at~~} y=0,~L_1\pm\epsilon,~ L)
\end{equation}
More specifically, we can take $\Psi_R=0$ (or $\Psi_L=0$) at all the boundary points to realize the left-handed (or right-handed) fermions in the zero mode sector, as we will discuss later.

Multiplying the operator $(i\Gamma^N\partial_N-M_F)$ on eq.~\eqref{eom} from the left gives
\begin{eqnarray}
(i\Gamma^N\partial_N-M_F)(i\Gamma^M\partial_M+M_F)\Psi=
\left(\begin{matrix}-DD^\dag+\partial_\mu\partial^\mu&\\
&-D^\dag D+\partial_\mu\partial^\mu\end{matrix}\right)
\left(\begin{matrix}\Psi_L\\\Psi_R\end{matrix}\right)=0 \label{SUSYQM-1}
\end{eqnarray}
where $D\equiv\partial_y+M_F$, $D^\dag\equiv-\partial_y+M_F$, and $\partial_\mu\partial^\mu\equiv\eta_{\mu\nu}\partial^\mu\partial^\nu=-\partial_t^2+\nabla^2$ with the 4D metric $\eta_{\mu\nu}=\mathrm{diag}(-1,1,1,1)$.
Let us separate variables of the solutions of eq.~\eqref{SUSYQM-1} as follows
\begin{equation}
\Psi_L(x,y)=\sum_n\psi_L^{(n)}(x)f_{\psi_L^{(n)}}(y)\;,\qquad
\Psi_R(x,y)=\sum_n\psi_R^{(n)}(x)f_{\psi_R^{(n)}}(y)\label{modes}
\end{equation}
For every particular solution of the left-handed wave-function, $\Psi_L(x,y)=\psi_L^{(n)}(x)f_{\psi_L^{(n)}}(y)$, we have
\begin{eqnarray}
0&=&(-DD^\dag+\partial_\mu\partial^\mu)\psi_L^{(n)}(x)f_{\psi_L^{(n)}}(y)\nonumber\\
&=&\left[-DD^\dag f_{\psi_L^{(n)}}(y)\right]\psi_L^{(n)}(x)
+\left[\partial_\mu\partial^\mu\psi_L^{(n)}(x)\right]f_{\psi_L^{(n)}}(y)\nonumber\\
&=&\left[\left(-DD^\dag+M_{\psi^{(n)}}^2 \right)f_{\psi_L^{(n)}}(y)\right]\psi_L^{(n)}(x)
\label{SUSYQM-2-LH}
\end{eqnarray}
where we have used the 4D Klein-Gordon equation $(\partial_\mu\partial^\mu-M_{\psi^{(n)}}^2)\psi_L^{(n)}(x)=0$. Eq.~\eqref{SUSYQM-2-LH} implies
\begin{subequations}
\begin{equation}
DD^\dag f_{\psi_L^{(n)}}(y)=M_{\psi^{(n)}}^2f_{\psi_L^{(n)}}(y) \label{SUSYQM-3-LH}
\end{equation}
Likewise, using $(\partial_\mu\partial^\mu-M_{\psi^{(n)}}^2)\psi_R^{(n)}(x)=0$, we obtain
\begin{eqnarray}
D^\dag Df_{\psi_R^{(n)}}(y)=M_{\psi^{(n)}}^2f_{\psi_R^{(n)}}(y) \label{SUSYQM-3-RH}
\end{eqnarray}
\end{subequations}
In eqs.~\eqref{SUSYQM-3-LH} and \eqref{SUSYQM-3-RH}, we have used the fact that the operators $DD^\dag$ and $D^\dag D$ are supersymmetric quantum mechanical partners \cite{Witten:1981nf,ArkaniHamed:1999dc,Cooper:1994eh} and thus they have exactly the same eigenvalues except for the lowest zero eigenvalue. It can be easily explained as follows.
If $f_{\psi_L^{(n)}}(y)$ is the eigenfunction of $DD^\dag$ with the eigenvalue $M_{\psi^{(n)}}^2$ and $M_{\psi^{(n)}}^2\neq 0$, then
\begin{equation}
D^\dag D\left[D^\dag f_{\psi_L^{(n)}}(y)\right]=D^\dag \left[DD^\dag f_{\psi_L^{(n)}}(y)\right]=M_{\psi^{(n)}}^2\left[D^\dag f_{\psi_L^{(n)}}(y)\right]
\end{equation}
that is, $D^\dag f_{\psi_L^{(n)}}(y)$ is an eigenfunction of $D^\dag D$ with the same eigenvalue $M_{\psi^{(n)}}^2$. Define $f_{\psi_R^{(n)}}(y)\propto D^\dag f_{\psi_L^{(n)}}(y)$ and let $f_{\psi_R^{(n)}}(y)$ have the same normalization as $f_{\psi_L^{(n)}}(y)$:
\begin{equation}
\langle f_{\psi_L^{(n)}}(y)|f_{\psi_L^{(n)}}(y)\rangle\equiv
\int dy\left[f_{\psi_L^{(n)}}(y)\right]^\ast f_{\psi_L^{(n)}}(y)=1 \label{normalization-condition}
\end{equation}
which implies
\begin{equation}
\int dy\left[D^\dag f_{\psi_L^{(n)}}(y)\right]^\ast D^\dag f_{\psi_L^{(n)}}(y)
=\int dy\left[f_{\psi_L^{(n)}}(y)\right]^\ast DD^\dag f_{\psi_L^{(n)}}(y)
=M_{\psi^{(n)}}^2
\end{equation}
Then it is sufficient to get $\langle f_{\psi_R^{(n)}}(y)|f_{\psi_R^{(n)}}(y)\rangle=1$ by letting
\begin{subequations}
\begin{equation}
f_{\psi_R^{(n)}}(y)=\frac{1}{M_{\psi^{(n)}}}D^\dag f_{\psi_L^{(n)}}(y) \label{fL-R-1}
\end{equation}
Multiplying the operator $D$ on the above equation from the left gives
\begin{equation}
f_{\psi_L^{(n)}}(y)=\frac{1}{M_{\psi^{(n)}}}D f_{\psi_R^{(n)}}(y) \label{fL-R-2}
\end{equation}
\end{subequations}

Substituting a pair of chiral modes of \eqref{modes}
into eq.~\eqref{eom},
\begin{equation}
\left(\begin{matrix}D^\dag &i\sigma^\mu\partial_\mu\\
i\bar{\sigma}^\mu\partial_\mu& D\end{matrix}\right)
\left(\begin{matrix}\psi_L^{(n)}(x)f_{\psi_L^{(n)}}(y) \\ \psi_R^{(n)}(x)f_{\psi_R^{(n)}}(y)\end{matrix}\right)=0
\end{equation}
we have
\begin{subequations}
\begin{eqnarray}
&&\psi_L^{(n)}(x)\left[D^\dag f_{\psi_L^{(n)}}(y)\right]
+\left[i\sigma^\mu\partial_\mu \psi_R^{(n)}(x)\right]f_{\psi_R^{(n)}}(y)=0\\
&&\left[i\bar{\sigma}^\mu\partial_\mu \psi_L^{(n)}(x)\right]f_{\psi_L^{(n)}}(y)
+\psi_R^{(n)}(x)\left[Df_{\psi_R^{(n)}}(y)\right]=0
\end{eqnarray}
\end{subequations}
which, together with eqs.~\eqref{fL-R-1} and \eqref{fL-R-2}, lead to
\begin{subequations}
\begin{eqnarray}
&&i\sigma^\mu\partial_\mu \psi_R^{(n)}(x)+M_{\psi^{(n)}}\psi_L^{(n)}(x)=0\\
&&i\bar{\sigma}^\mu\partial_\mu \psi_L^{(n)}(x)++M_{\psi^{(n)}}\psi_R^{(n)}(x)=0
\end{eqnarray}
\end{subequations}
that is,
\begin{equation}
\left(\begin{matrix}M_{\psi^{(n)}} &i\sigma^\mu\partial_\mu\\
i\bar{\sigma}^\mu\partial_\mu& M_{\psi^{(n)}}\end{matrix}\right)
\left(\begin{matrix}\psi_L^{(n)}(x) \\ \psi_R^{(n)}(x)\end{matrix}\right)=0
\end{equation}
Thus, the combination $\psi^{(n)}(x)\equiv \left(\psi_L^{(n)}(x),~\psi_R^{(n)}(x)\right)^T$ obeys the 4D Dirac equation $\left(i\slashed{\partial}+M_{\psi^{(n)}}\right)\psi^{(n)}(x)=0$ and forms a Dirac spinor.

\begin{figure}[!htbp]
\begin{center}
     \includegraphics[width=0.7\textwidth]{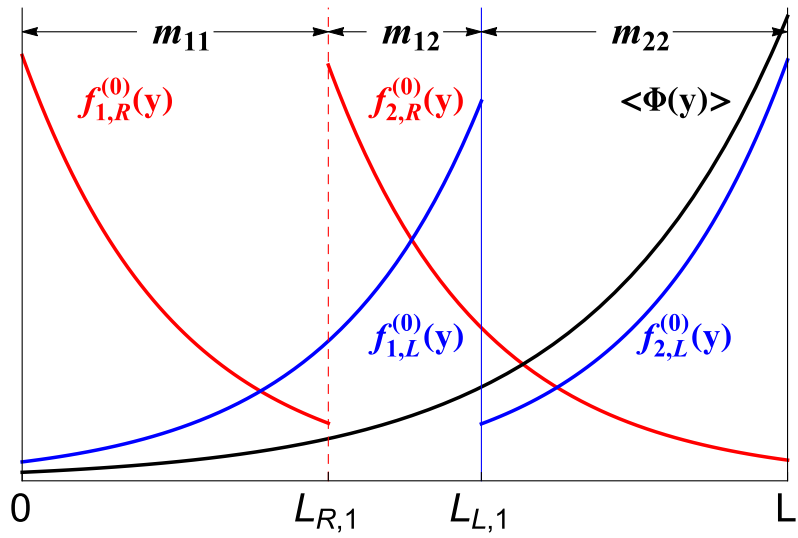}
\end{center}\caption{A schematic diagram of wave functions for chiral 0-mode fermions. The red curves represent the wave functions for two generations of right-handed 0-mode fermions, while the blue curves represent the wave functions for two generations of left-handed 0-mode fermions. The black line is a profile of a scalar $\Phi$'s VEV. The overlap integration of the profiles in different interval gives the corresponding mass matrix element.}\label{profiles}
\end{figure}

Suppose that the eigenequation \eqref{SUSYQM-3-LH} of $DD^\dag$ has a zero eigenvalue $M_{\psi^{(n)}}^2=0$ with the corresponding eigenfunction $f_{\psi_L^{(0)}}(y)$ called the 0-mode. That is, $DD^\dag f_{\psi_L^{(0)}}(y)=0$. It is sufficient to satisfy the above relation if $f_{\psi_L^{(0)}}(y)$ is annihilated by $D^\dag$:
\begin{equation}
D^\dag f_{\psi_L^{(0)}}(y)=(-\partial_y+M_F)f_{\psi_L^{(0)}}(y)=0 \label{left-hand-0-mode-eq}
\end{equation}
\begin{itemize}
  \item If the Dirichlet boundary condition $\Psi_L(x,y)=0$ is imposed at $y=0,~L_1\pm\epsilon,~L$, that is,
  \begin{equation}
  f_{\psi_L^{(0)}}(y)=0\qquad (\text{at~~} y=0,~L_1\pm\epsilon,~ L) \label{boundary-LH-zero}
  \end{equation}
  then eqs.~\eqref{left-hand-0-mode-eq} and \eqref{boundary-LH-zero} imply that $f_{\psi_L^{(0)}}(y)=0$ at all points. Thus, the 0-mode eigenfunction of $DD^\dag$ does not exist in the boundary condition of \eqref{boundary-LH-zero}.

  \item If the Dirichlet boundary condition $\Psi_R(x,y)=0$ is imposed at $y=0,~L_1\pm\epsilon,~L$, that is,
  \begin{equation}
  f_{\psi_R^{(0)}}(y)=0\qquad (\text{at~~} y=0,~L_1\pm\epsilon,~ L) \label{boundary-RH-zero}
  \end{equation}
  then this boundary condition has no effect on the equation \eqref{left-hand-0-mode-eq}, but the setup of the 0-thickness branes' positions itself can split the solutons of \eqref{left-hand-0-mode-eq} into two independent degenerate modes:
  \begin{subequations}
  \begin{eqnarray}
  &&f_{\psi_L^{(0)},(1)}(y)=\left\{\begin{array}{ll}
                               N_1e^{M_Fy} & \quad(0\leq y <L_1) \\
                               0 & \quad(L_1\leq y <L)
                             \end{array}\right.\\
  &&f_{\psi_L^{(0)},(2)}(y)=\left\{\begin{array}{ll}
                               0 & \quad(0\leq y <L_1) \\
                               N_2e^{M_Fy} & \quad(L_1\leq y <L)
                             \end{array}\right.
  \end{eqnarray}
  \end{subequations}
  where $N_1$ and $N_2$ are normalization constants and, by using \eqref{normalization-condition}, they can be figured out as
  \begin{equation}
  N_1=\sqrt{\frac{2M_F}{e^{2M_FL_1}-1}}\;,\qquad
  N_2=e^{-M_FL_1}\sqrt{\frac{2M_F}{e^{2M_F(L-L_1)}-1}}
  \end{equation}
  Using the Heaviside step function $\theta(y)$, we can also write the two degenerate zero modes as follows
  \begin{subequations}
  \begin{eqnarray}
  &&f_{\psi_L^{(0)},(1)}(y)=\sqrt{\frac{2M_F}{e^{2M_FL_1}-1}}e^{M_Fy}
  \left[\theta(y)\theta(L_1-y)\right]\\
  &&f_{\psi_L^{(0)},(2)}(y)=\sqrt{\frac{2M_F}{e^{2M_F(L-L_1)}-1}}e^{M_F(y-L_1)}
  \left[\theta(y-L_1)\theta(L-y)\right]
  \end{eqnarray}
  \end{subequations}
  The 5D wavefunction of 0-mode $\Psi_L^{(0)}(x,y)$ may be expanded with respect to $f_{\psi_L^{(0)},(1)}(y)$ and $f_{\psi_L^{(0)},(2)}(y)$ as
  \begin{eqnarray}
  \Psi_L^{(0)}(x,y)=\psi_{1L}^{(0)}(x)f_{\psi_L^{(0)},(1)}(y)
  +\psi_{2L}^{(0)}(x)f_{\psi_L^{(0)},(2)}(y)\label{left-hand-0-mode-eq-1}
  \end{eqnarray}
  where the coefficients $\psi_{1L}^{(0)}(x)$ and $\psi_{2L}^{(0)}(x)$ are identified with the 4D wavefunctions of two generations of left-handed fermions in this toy model.
\end{itemize}
Likewise, consider the 0-mode eigenfunction $f_{\psi_R^{(0)}}(y)$ of $D^\dag D$. It obeys the equation $D^\dag Df_{\psi_R^{(0)}}(y)=0$. A sufficient condition of this equation is
\begin{equation}
Df_{\psi_R^{(0)}}(y)=(\partial_y+M_F)f_{\psi_R^{(0)}}(y)=0
\label{right-hand-0-mode-eq}
\end{equation}
\begin{itemize}
  \item If the Dirichlet boundary condition \eqref{boundary-LH-zero} for the left-handed fermion is imposed, then it is the location of the point-interaction positions, rather than eq.~\eqref{boundary-LH-zero}, that affects the solutions of \eqref{right-hand-0-mode-eq} and splits them into two degenerate modes:
  \begin{subequations}
  \begin{eqnarray}
  &&f_{\psi_R^{(0)},(1)}(y)
  =\sqrt{\frac{2M_F}{1-e^{-2M_FL_1}}}e^{-M_Fy}[\theta(y)\theta(L_1-y)]\\
  &&f_{\psi_R^{(0)},(2)}(y)
  =\sqrt{\frac{2M_F}{1-e^{-2M_F(L-L_1)}}}e^{-M_F(y-L_1)}[\theta(y-L_1)\theta(L-y)]
  \end{eqnarray}
  \end{subequations}
  The expansion of the 5D wavefunction of 0-mode $\Psi_R^{(0)}(x,y)$ with respect to the two modes is given by
  \begin{eqnarray}
  \Psi_R^{(0)}(x,y)=\psi_{1R}^{(0)}(x)f_{\psi_R^{(0)},(1)}(y)
  +\psi_{2R}^{(0)}(x)f_{\psi_R^{(0)},(2)}(y) \label{right-hand-0-mode-eq-1}
  \end{eqnarray}
  where the 4D wavefunctions $\psi_{1R}^{(0)}(x)$ and $\psi_{2R}^{(0)}(x)$ belong to two generations of right-handed fermions in this toy model.

  \item If the Dirichlet boundary condition \eqref{boundary-RH-zero} for the right-handed fermion is imposed, then eqs.~\eqref{right-hand-0-mode-eq} and \eqref{boundary-RH-zero} imply that $f_{\psi_R^{(0)}}(y)=0$ at all points. That is, the 0-mode eigenfunction of $D^\dag D$ vanishes in this boundary condition.
\end{itemize}
To sum up, if the boundary condition $\Psi_L=0$ is imposed at all the 0-thickness branes' positions, then the 5D fermion field $\Psi(x,y)$ has only right-handed 0-modes $\Psi_R^{(0)}(x,y)$ as given in eq.~\eqref{right-hand-0-mode-eq-1}; instead, if $\Psi_R=0$ is imposed at all the boundary points, then $\Psi(x,y)$ has only left-handed 0-modes $\Psi_L^{(0)}(x,y)$ as given in eq.~\eqref{left-hand-0-mode-eq-1}. In a word, the Dirichlet boundary condition $\Psi_{L,R}=0$ makes the 0-mode wavefunctions of $\Psi(x,y)$ to be chiral. Including the KK modes (\textit{i.e.} the modes with $M_{\psi^{(n)}}^2\neq 0$), the expansion of a 5D fermion field $\Psi(x,y)$ in all modes is given by
\begin{itemize}
\item For $\Psi_L=0$ at $y=0,L_1,L$
\begin{eqnarray}
\Psi(x,y)&=&\sqrt{\frac{2M_F}{1-e^{-2M_FL_1}}}e^{-M_Fy}[\theta(y)\theta(L_1-y)]\psi_{1R}^{(0)}(x)\nonumber\\
&&+\sqrt{\frac{2M_F}{1-e^{-2M_F(L-L_1)}}}e^{-M_F(y-L_1)}[\theta(y-L_1)\theta(L-y)]\psi_{2R}^{(0)}(x)\nonumber\\
&&+\,\mathrm{(KK~modes)}
\end{eqnarray}
\item For $\Psi_R=0$ at $y=0,L_1,L$
\begin{eqnarray}
\Psi(x,y)&=&\sqrt{\frac{2M_F}{e^{2M_FL_1}-1}}e^{M_Fy}[\theta(y)\theta(L_1-y)]\psi_{1L}^{(0)}(x)\nonumber\\
&&+\sqrt{\frac{2M_F}{e^{2M_F(L-L_1)}-1}}e^{M_F(y-L_1)}[\theta(y-L_1)\theta(L-y)]\psi_{2L}^{(0)}(x)\nonumber\\
&&+\,\mathrm{(KK~modes)}
\end{eqnarray}
\end{itemize}
To realize both left-handed and right-handed 0-mode fermions in this 2-generation toy model, we need at least two 5D fermion fields, $\Psi_1(x,y)$ and $\Psi_2(x,y)$. One 5D fermion $\Psi_1(x,y)$ has two left-handed 0-modes due to the boundary condition $P_R\Psi_{1}(x,y)=0$ at points $y=0,~L_1,~L$; while another 5D fermion $\Psi_2(x,y)$ has two right-handed 0-modes from the boundary condition $P_L\Psi_{2}(x,y)=0$ at points $y=0,~L_1^\prime,~L$. The locations of $L_1$ and $L_1^\prime$ are in general not equal. Indeed, it is the inequality of $L_1$ and $L_1^\prime$ that leads to the mixing of the two generations of fermions. A schematic picture of the wave functions of these 0-mode chiral fermions is shown in Fig.~\ref{profiles}.
To give the chiral fermions masses, we need to introduce an extra 5D scalar field $\Phi(x,y)$, which will acquire a nonzero VEV after the electroweak symmetry breaking. The mixing structure of the Dirac mass matrix is also explained in Fig.~\ref{profiles}.

In addition, it is worthy to point out that the operators $D\equiv\partial_y+M_F$ and $D^\dag\equiv-\partial_y+M_F$ can be used to construct a pair of supersymmetric generators, $Q\equiv D\gamma^0P_L$ and $Q^\dag\equiv D^\dag\gamma^0P_R$, which satisfy the supersymmetric algebra (See the paragraphs between eqs.~(8) and (9) in Ref.~\cite{ArkaniHamed:1999dc} for more details):
\begin{eqnarray}
Q^2=Q^{\dag2}=0\,,\quad
\{Q,Q^\dag\}=2H\,,\quad
[Q,H]=[Q^\dag,H]=0
\end{eqnarray}
The Hamiltonian operators (up to a constant factor) is $H\propto\{Q,Q^\dag\}=DD^\dag P_R+D^\dag DP_L$, and the pair of modes $(f_{\psi_L^{(n)}}(y)\psi_L^{(n)}(x),~f_{\psi_R^{(n)}}(y)\psi_R^{(n)}(x))^T$ is an eigenstate of $H$ with eigenvalue $M_{\psi^{(n)}}^2$.
\section{Quark masses hierarchy and flavor mixings}\label{app2}
The Yukawa terms which generate the masses for quarks are:
\begin{eqnarray}\label{quarkyukawa}
  \mathcal{L}^{Yuk}_{quarks}=-\int dy[\mathcal{Y}^{(u)}\Phi\overline{Q}(i\sigma^2H^\ast)U_R+\mathcal{Y}^{(d)}\Phi^\ast\overline{L}HD_R+h.c.]
\end{eqnarray}
where $\mathcal{Y}^{(u)}$ and $\mathcal{Y}^{(d)}$ are the couplings with dimension $-2$ for the up type and down type quarks, respectively.

Note that we will let $\Phi$ to be $U(1)'$ charged. Then if we don't want the $U(1)'$ breaks explicitly, we should also make $U_R$, $D_R$, $Q$ and $H$ to be $U(1)'$ charged.
We have determined the $U(1)'$ charge for each field in section \ref{sect3}. We can see that terms as $\overline{Q}(i\sigma^2H^\ast)U_R$ and $\overline{L}HD_R$ can be forbidden by the $U(1)'$ symmetry.\par
After the $U(1)'$ and electro-weak symmetry breaking,we obtain Dirac mass terms of quarks.
The mixing structure of the mass matrix will be generated by the overlaps of wave functions from different generations.
Then we can write down the mass matrices as
\begin{eqnarray}
&&m^{(u)}=\left(\begin{matrix}&m^{u}_{11}&m^{u}_{12}&m^{u}_{13}\\&0&m^{u}_{2}&m^{u}_{23}\\&0&0&m^{u}_{33}\end{matrix}\right),\ m^{(d)}=\left(\begin{matrix}&m^{d}_{11}&m^{d}_{12}&m^{d}_{13}\\&0&m^{d}_{2}&m^{d}_{23}\\&0&0&m^{d}_{33}\end{matrix}\right)\\
&&m^{(u)}_{ij}=\mathcal{Y}^{(u)}\int_a^bdyf_{q^{(0)}_{iL}}(y)f_{u^{(0)}_{jR}}(y)\langle\phi(y)\rangle\langle H(y)^\ast\rangle\\
&&m^{(d)}_{ij}=\mathcal{Y}^{(d)}\int_a^bdyf_{q^{(0)}_{iL}}(y)f_{d^{(0)}_{jR}}(y)\langle\phi(y)\rangle\langle H(y)\rangle
\end{eqnarray}
The integration range $(a,b)$ represents the overlap region between the profiles $f_{q^{(0)}_{iL}}(y)$ and $f_{u^{(0)}_{jL}}(y)$ or $f_{d^{(0)}_{jL}}(y)$.
The integration will contribute to a diagonal element when $i=j$, and an off diagonal element when $i\neq j$.
Two Dirac mass matrices $m^{(u)}$ and $m^{(d)}$ are apparently complex and we can diagonalize them with unitary matrices $V_L^{(u)}(V_L^{(d)})$ and $V_R^{(u)}(V_R^{(d)})$.
\begin{eqnarray}
\left\{\begin{array}{l}
m_{diag}^{(u)}=V_L^{(u)}m^{(u)}V_R^{(u)\dag}\\
m_{diag}^{(d)}=V_L^{(d)}m^{(d)}V_R^{(d)\dag}
\end{array}\right.
\end{eqnarray}
Then we can compare the masses with experimental data.
Using matrices $V_L^{(u)}$ and $V_L^{(d)}$, we can calculate the CKM matrix which is defined as
\begin{eqnarray}
V_{CKM}&=&V_L^{(u)}V_L^{(d)\dag}
\end{eqnarray}
The CKM matrix contains not only information about flavor mixing angles but also information about the CP violation.
The CP violation can be characterized by the Jarlskog invariant $\mathcal{J}$ defined as
\begin{eqnarray}
\mathrm{Im}[(V_{CKM})_{ij}(V_{CKM})_{kl}(V_{CKM}^\ast)_{il}(V_{CKM}^\ast)_{kj}]=\mathcal{J}\sum_{m,n=1}^3\epsilon_{ikm}\epsilon_{jln}
\end{eqnarray}
We list the experimental data used in our fitting as follows
\begin{itemize}
\item The up and down type quark masses are shown in Table.~\ref{quarkmass}
\begin{table}[!htbp]
\caption{Quark masses from ref.\cite{Beringer:1900zz}}\label{quarkmass}
\centering
\begin{tabular}{|c|c|c|c|}
\hline
up type quark    & mass                         & down type quark     &mass\\
\hline
u                         & $2.3\pm0.6$ MeV         & d                              &$4.8\pm0.5$ MeV\\
\hline
c                         & $1.275\pm0.025$ GeV& s                              &$95\pm5$ MeV\\
\hline
t                         & $173.5\pm1.4$ GeV     & b                              &$4.18\pm0.03$ GeV\\
\hline
\end{tabular}
\end{table}
\item The absolute values of CKM matrix elements from ref. \cite{Beringer:1900zz} are
\begin{eqnarray}
|V_{CKM}|=\left(\begin{matrix}0.97425\pm0.00022&0.2252\pm0.0009&0.00415\pm0.049\\0.230\pm0.011&1.006\pm0.023&0.0409\pm0.0011\\0.0084\pm0.0006&0.0429\pm0.0026&0.89\pm0.07\end{matrix}\right)
\end{eqnarray}
\item The Jarlskog invariant from ref.\cite{Beringer:1900zz} is $\mathcal{J}=(2.96\pm0.18)\times10^{-5}$.
\end{itemize}
After fitting the data listed above. We find a set of parameters, which is compatible with the data, and show them in Table.~\ref{quarkbf}.
\begin{table}[!htbp]
\renewcommand\arraystretch{1.25}
\caption{Best fit parameters for quarks}\label{quarkbf}
\centering
\begin{tabular}{|c|c|c|c|}
\hline
$L^{(q)}_0$    &$L^{(q)}_1$  &$L^{(q)}_2$  & $M_Q$\\
\hline
$0$            &$0.31423L$   &$0.67665L$   &$9.26018L^{-1}$\\
\hline
\hline
$L^{(u)}_0$    &$L^{(u)}_1$  &$L^{(u)}_2$  & $M_U$\\
\hline
$0.05218L$     &$0.06095L$   &$0.56328L$   &$-4.48152L^{-1}$\\
\hline
\hline
$L^{(d)}_0$    &$L^{(d)}_1$  &$L^{(d)}_2$  & $M_D$\\
\hline
$0.11866L$     &$0.23128L$   &$0.66636L$   &$5.71010L^{-1}$\\
\hline
\hline
$M$            &$\tilde{\mathcal{Y}}^{(u)}v/\sqrt{2}$    &$\tilde{\mathcal{Y}}^{(d)}v/\sqrt{2}$    & $\theta$\\
\hline
$9.36099L^{-1}$&$3.15684$ GeV     &$0.20552$ GeV     &2.91684\\
\hline
\end{tabular}
\end{table}
We have set $|\tilde{\lambda}|\equiv|\lambda L|=0.001,|\tilde{Q}|\equiv|QL^5|=0.001$ and $\tilde{y}_0\equiv y_0L^{-1}=-0.16$ fixed as reference \cite{Fujimoto:2012wv} did, so the only free parameter of $\Phi$ is $M$. Since $\Phi$ and $H$ also couple to leptons,the values of $M$ and $\theta$ which are found in the quark case will be set fixed to reduce the number of free parameters in the lepton case. In the following,a parameter with a tilde means it has been scaled to dimensionless by multiply some power of $L$.\par

Note that we can calculate $L_{\pm}$ in the Robin boundary condition by:
\begin{eqnarray}
\left\{\begin{array}{l}
L+=-\frac{\Phi(0)}{\partial_y\Phi(0)}=-0.118074L\\
L_-=\frac{\Phi(L)}{\partial_y\Phi(L)}=0.104502L\end{array}\right.
\end{eqnarray}
Then we find that $M=9.36099<\frac{1}{L_-}=9.5692$, which is consistent with the symmetry breaking condition $|M|^2<\frac{1}{L_{max}^2}$.\par
Using the parameters of $\Phi$ we can calculate the tree level mass of the 4D excitation $\phi(x)$.
One of its degree of freedom will be gauged out by the gauge boson of $U(1)'$ when the symmetry breaking occurs.
To obtain the mass of $\phi(x)$, we shall consider its excitation around the minimum of potential
\begin{eqnarray}
\mathcal{E}[\Phi]=\int_0^Ldy\{-\Phi^\dag\partial_y^2\Phi+M^2|\Phi|^2+\frac{\lambda}{2}|\Phi|^4\}
\end{eqnarray}
Substitute the zero mode $\Phi^{(0)}=f^{(0)}(y)(\nu+\phi),\nu f^{(0)}(y)=\langle\Phi(y)\rangle$ into $\mathcal{E}[\Phi]$ and use the minimized condition:$-\partial_y^2f_0(y)+M^2f_0+\lambda\nu^2f_0^3=0$, we can get the mass
\begin{eqnarray}
m_\phi^2=2\int_0^Ldy(2\lambda\langle\Phi(y)\rangle^2f_0^2)
\approx\frac{\lambda|Q|}{M^2}e^{2M(L-y_0)}
=\frac{\tilde{\lambda}|\tilde{Q}|}{\tilde{M}^2}e^{2\tilde{M}(1-\tilde{y}_0)}L^{-2}
\end{eqnarray}
which implies
\begin{eqnarray}
m_\phi&\approx&\frac{\sqrt{\tilde{\lambda}|\tilde{Q}|}}{\tilde{M}}e^{\tilde{M}(1-\tilde{y}_0)}L^{-1}\approx5.55\textrm{TeV}\cdot\left(\frac{L^{-1}}{\textrm{TeV}}\right)
\end{eqnarray}
If the scale $L^{-1}\sim$O(1TeV), this mass is under the energy scale of LHC. But it is unlikely to be detected in the recent experiments, because the the $\phi$-fermion-fermion couplings are so weak.
This can be seen by estimate the couplings as
\begin{eqnarray}
\zeta^{(q)}_{ij}&=&\frac{m^{(q)}_{ij}\cdot A}{\nu}\,,\qquad \zeta^{(e)}_{ij}=\frac{m^{(e)}_{ij}\cdot A}{\nu}
\end{eqnarray}
where
\begin{eqnarray}
&&A=\frac{\sqrt{2/L}}{\sqrt{1+\sinh(\tilde{M})\cosh(\tilde{M}-2\tilde{M}\tilde{y}_0)/\tilde{M}}}
\simeq\sqrt{\frac{2}{L}}\frac{2\sqrt{\tilde{M}}}{e^{\tilde{M}(1-\tilde{y}_0)}}\\
&&\nu\simeq\frac{\sqrt{2|\tilde{Q}|}}{\tilde{M}}L^{-\frac{3}{2}}
\end{eqnarray}
Using the parameters in our fitting, we find the Yukawa couplings for $\phi$-quark-quark and $\phi$-lepton-lepton are
\begin{eqnarray}
\zeta^{(q)}_{ij}\simeq0.03\times\frac{m^{(q)}_{ij}}{L^{-1}}\,,\qquad
\zeta^{(e)}_{ij}\simeq0.03\times\frac{m^{(e)}_{ij}}{L^{-1}}
\end{eqnarray}
Both Yukawa couplings are much weaker than the Yukawa couplings for Higgs-quark-quark and Higgs-lepton-lepton.
Since the coupling is proportional to the mass, the strongest Yukawa coupling may be the coupling of $\phi$-top-top which is about $0.03\times0.17\approx0.005$ when $L^{-1}\sim 1$ TeV.\par
Note that there is a $C|\Phi|^2|H|^2$ term may lead to some problem with the gauge universality as discussed in reference \cite{Fujimoto:2012wv}.
We will just let $C$ to be small enough (about $10^{-7}$ for $L^{-1}\sim1$ TeV) to resolve this.
\section{Why an explicit Majorana mass term does not work}\label{app3}
The 5D charge conjugation operator $C$ is defined as
\begin{eqnarray}
C\Gamma^MC^{-1}=(\Gamma^M)^T
\end{eqnarray}
with properties:
\begin{eqnarray}
C^T=C^{-1}=C^\dag=-C
\end{eqnarray}
It is easy to check that $C$ can be written as $C=\gamma^0\gamma^2(i\gamma_5)$ \cite{ArkaniHamed:1999dc}.
We can write it in Weyl basis
\begin{eqnarray}
C=\begin{pmatrix}\epsilon_{ab}&\\&-\epsilon^{\dot{a}\dot{b}}\end{pmatrix}
\end{eqnarray}
The charge conjugation of a 5D fermion is defined as
\begin{eqnarray}
\Psi^c=C\bar{\Psi}^T
\end{eqnarray}
We can also write it down in Weyl basis:
\begin{eqnarray}
\Psi(x,y)=\begin{pmatrix}\xi_a(x,y)\\ \chi^{\dag\dot{a}}(x,y)\end{pmatrix}\Rightarrow\Psi^c=\begin{pmatrix}\chi_a(x,y)\\-\xi^{\dag\dot{a}}(x,y)\end{pmatrix}
\end{eqnarray}
Note that the relation $(\Psi^c)^c=\Psi$ no longer holds in 5D case and the correct relation is $(\Psi^c)^c=-\Psi$.\par
Now we consider to add terms as $\bar{\Psi}i\Gamma^M\partial_M\Psi^c+h.c.$, after several lines of calculation, we can get
\begin{eqnarray}
\bar{\Psi}i\Gamma^M\partial_M\Psi^c&=&\bar{\Psi}i\Gamma^M\partial_MC\bar{\Psi}^T\nonumber\\
&=&\partial_M(\bar{\Psi}i\Gamma^M\Psi^c)-\bar{\Psi}i\Gamma^M\partial_M\Psi^c
\end{eqnarray}
This implies that these terms can be absorbed into the boundary terms and do not contribute to the equations of motion.\par
However, the mass terms as $M_R\bar{\Psi}\Psi^c+h.c.$ survive and will contribute to the equations of motion.
Now let's add the mass terms into the action:
\begin{eqnarray}\label{act2}
S=\int d^4x\int dy[\bar{\Psi}(x,y)(i\Gamma^M\partial_M+M_F)\Psi(x,y)+\frac{1}{2}(M_R\bar{\Psi}\Psi^c+h.c.)]
\end{eqnarray}
The variation of the action \eqref{act2} is:
\begin{eqnarray}\label{variation-1}
\delta S&=&\int d^4x\int dy\left[\delta\bar{\Psi}(i\Gamma^M\partial_M+M_F)\Psi+\bar{\Psi}(i\Gamma^M\partial_M+M_F)\delta\Psi\phantom{\frac{1}{2}}\right.\nonumber\\
&&\left.\phantom{\int d^4x\int dy}+\frac{1}{2}M_R\delta\bar{\Psi}\Psi^c+\frac{1}{2}M_R\bar{\Psi}\delta\Psi^c+\frac{1}{2}M_R\delta\overline{\Psi^c}\Psi+\frac{1}{2}M_R\overline{\Psi^c}\delta\Psi\right]\nonumber\\
&=&\int d^4x\int dy\left[\delta\bar{\Psi}(i\Gamma^M\partial_M+M_F)\Psi
-\bar{\Psi}(i\Gamma^M\overleftarrow{\partial}_M-M_F)\delta\Psi\right.\nonumber\\
&&\left.\phantom{\int d^4x\int dy}+\partial_M(\bar{\Psi}i\Gamma^M\delta\Psi)+M_R\delta\bar{\Psi}\Psi^c+M_R\overline{\Psi^c}\delta\Psi\right]\nonumber\\
\end{eqnarray}
Thus, the equation of motion (EOM) becomes:
\begin{eqnarray}\label{eom2}
0&=&(i\Gamma^M\partial_M+M_F)\Psi+M_R\Psi^c\nonumber\\
&=&\left(\begin{matrix}-\partial_y+M_F&i\sigma^\mu\partial_\mu\\
i\bar{\sigma}^\mu\partial_\mu&\partial_y+M_F\end{matrix}\right)
\begin{pmatrix}\xi_a(x,y)\\ \chi^{\dag\dot{a}}(x,y)\end{pmatrix}+\begin{pmatrix}M_R&\\&M_R\end{pmatrix}\begin{pmatrix}\chi_a(x,y)\\-\xi^{\dag\dot{a}}(x,y)\end{pmatrix}
\end{eqnarray}
If we try to separate the field in modes as $\xi_a(x,y)=\sum_nf^{(n)}(y)\xi^{(n)}_a(x),\chi_a=\sum_ng^{(n)}(y)\chi^{(n)}_a(x)$, then the equations for each mode become
\begin{eqnarray}
(-\partial_y+M_F)f^{(n)}(y)\xi^{(n)}_a(x)+M_Rg^{(n)}(y)\chi^{(n)}_a(x)+g^{(n)\ast}(y)i\sigma^\mu\partial_\mu\chi^{(n)\dag\dot{a}}(x)=0\\
(\partial_y+M_F)g^{(n)\ast}(y)\chi^{(n)\dag\dot{a}}(x)-M_Rf^{(n)\ast}(y)\xi^{(n)\dag\dot{a}}(x)+f^{(n)}i\bar{\sigma}^\mu\partial_\mu\xi^{(n)}_a(x)=0
\end{eqnarray}
Apparently, in a general case, it is impossible to factor out the functions $f^{(n)}(y),g^{(n)}(y)$ from the 4D Dirac equations of spinors $\xi_a(x),\chi_a(x)$. This means A special choice which can achieve this is to let $M_F=0$ and $\chi^{\dag\dot{a}}=-\xi^{\dag\dot{a}}$, then the EOM become:
\begin{eqnarray}
(\partial_y+M_R)\xi_a(x,y)+i\sigma^\mu_{a\dot{a}}\partial_\mu\xi^{\dag\dot{a}}(x,y)=0\\
(-\partial_y+M_R)\xi^{\dag\dot{a}}(x,y)+i\bar{\sigma}^{\mu \dot{a}a}\partial_\mu\xi_a(x,y)=0
\end{eqnarray}
We can recover the 4D Dirac equation for a Majorana fermion by setting $\xi_a(x,y)=A\xi_a(x)$ where $A$ is a constant so the profile is independent of the 5th dimension coordinate $y$. Thus, this fermion has only one mode with a Majorana mass $M_R$. But this solution requires some special choice of the 5D fermion.\par
If we accept this special pattern of fermion to be the singlet neutrino $N_R$, and generate Dirac masses with the Yukawa interaction, then the seesaw turns on when the Majorana mass is much larger than the Dirac ones. However, an operator as $\overline{L}\sigma^2H^\ast H^\dag\sigma^2L^c$ is still allowed and it will contribute to the Majorana masses of left-handed zero modes. Now we have to diagonalize the following mass matrix
\begin{eqnarray}
\mathcal{M}=\begin{pmatrix}M_L&M_D\\M_D^T&M_R\end{pmatrix}
\end{eqnarray}
In the large $M_R$ limit, the light neutrino masses are $m_\nu\approx M_L-M_DM_D^T/M_R$.
These masses should be as small as O(0.1 eV) to fit the current neutrino mass bound and imply that either we use an unnaturally small coupling for $\overline{L}\sigma^2H^\ast H^\dag\sigma^2L^c$ operator or we fine-tuned the parameters to cancel $M_L$ by $M_DM_D^T/M_R$ in high precision.\par
Actually, in the SM the gauge symmetries and the lepton number conservation do not allow the explicit Majorana mass term and $\overline{L}\sigma^2H^\ast H^\dag\sigma^2L^c$ to exist.
However, in this model we are going to add a SM-gauge-group singlet neutrino field into the model, and try to violate the lepton number explicitly.  Thus, we have to face these annoying terms unless they are also forbidden by some symmetry. The strategy we use in the paper is to forbid both $\overline{L}\sigma^2H^\ast H^\dag\sigma^2L^c$ and $M_R\bar{\Psi}\Psi^c+h.c.$ terms by a U(1)' symmetry. Then the singlet neutrinos have chiral zero-modes as any other fermions. Their right-handed Majorana masses are generated by the VEV of $\Phi$ with the same mechanism as their Dirac masses generated by the VEV of $\Phi$ and the Higgs field. Now the mass matrix we need to diagonalize is
\begin{eqnarray}
\mathcal{M}=\begin{pmatrix}0&M_D\\M_D^T&M_R\end{pmatrix}
\end{eqnarray}
In large $M_R$ limit, the light neutrinos masses are $m_\nu\approx M_DM_D^T/M_R$ which can be naturally suppressed to O(0.1 eV).\par


\begin{thebibliography}{9}
\bibitem{Aad:2012tfa}
  G.~Aad {\it et al.}  [ATLAS Collaboration],
  ``Observation of a new particle in the search for the Standard Model Higgs boson with the ATLAS detector at the LHC,''
  Phys.\ Lett.\ B {\bf 716}, 1 (2012)
  [arXiv:1207.7214 [hep-ex]].
\bibitem{Chatrchyan:2012ufa}
  S.~Chatrchyan {\it et al.}  [CMS Collaboration],
  ``Observation of a new boson at a mass of 125 GeV with the CMS experiment at the LHC,''
  Phys.\ Lett.\ B {\bf 716}, 30 (2012)
  [arXiv:1207.7235 [hep-ex]].
\bibitem{Ade:2013zuv}
  P.~A.~R.~Ade {\it et al.}  [Planck Collaboration],
  ``Planck 2013 results. XVI. Cosmological parameters,''
  Astron.\ Astrophys.\  {\bf 571}, A16 (2014)
  [arXiv:1303.5076 [astro-ph.CO]].
\bibitem{Minkowski:1977sc}
  P.~Minkowski,
  ``$\mu \to e\gamma$ at a Rate of One Out of $10^{9}$ Muon Decays?,''
  Phys.\ Lett.\ B {\bf 67}, 421 (1977).
\bibitem{Yanagida:1979as}
  T.~Yanagida,
  ``Horizontal Symmetry And Masses Of Neutrinos,''
  Conf.\ Proc.\ C {\bf 7902131}, 95 (1979).
\bibitem{GellMann:1980vs}
  M.~Gell-Mann, P.~Ramond and R.~Slansky,
  ``Complex Spinors and Unified Theories,''
  Conf.\ Proc.\ C {\bf 790927}, 315 (1979)
  [arXiv:1306.4669 [hep-th]].
\bibitem{Mohapatra:1979ia}
  R.~N.~Mohapatra and G.~Senjanovic,
  ``Neutrino Mass and Spontaneous Parity Violation,''
  Phys.\ Rev.\ Lett.\  {\bf 44}, 912 (1980).
\bibitem{Schechter:1980gr}
  J.~Schechter and J.~W.~F.~Valle,
  ``Neutrino Masses in SU(2) x U(1) Theories,''
  Phys.\ Rev.\ D {\bf 22}, 2227 (1980).
\bibitem{Schechter:1981cv}
  J.~Schechter and J.~W.~F.~Valle,
  ``Neutrino Decay and Spontaneous Violation of Lepton Number,''
  Phys.\ Rev.\ D {\bf 25}, 774 (1982).
\bibitem{Foot:1988aq}
  R.~Foot, H.~Lew, X.~G.~He and G.~C.~Joshi,
  ``Seesaw Neutrino Masses Induced by a Triplet of Leptons,''
  Z.\ Phys.\ C {\bf 44}, 441 (1989).
\bibitem{ArkaniHamed:1999dc}
  N.~Arkani-Hamed and M.~Schmaltz,
  ``Hierarchies without symmetries from extra dimensions,''
  Phys.\ Rev.\ D {\bf 61} (2000) 033005
  [hep-ph/9903417].

\bibitem{Randall:1999ee}
  L.~Randall and R.~Sundrum,
  ``A Large mass hierarchy from a small extra dimension,''
  Phys.\ Rev.\ Lett.\  {\bf 83}, 3370 (1999)
  [hep-ph/9905221].

\bibitem{Grossman:1999ra}
  Y.~Grossman and M.~Neubert,
  ``Neutrino masses and mixings in nonfactorizable geometry,''
  Phys.\ Lett.\ B {\bf 474}, 361 (2000)
  [hep-ph/9912408].

\bibitem{Fujimoto:2012wv}
  Y.~Fujimoto, T.~Nagasawa, K.~Nishiwaki and M.~Sakamoto,
  ``Quark mass hierarchy and mixing via geometry of extra dimension with point interactions,''
  PTEP {\bf 2013} (2013) 023B07
  [arXiv:1209.5150 [hep-ph]].

\bibitem{Fujimoto:2014fka}
  Y.~Fujimoto, K.~Nishiwaki, M.~Sakamoto and R.~Takahashi,
  ``Realization of lepton masses and mixing angles from point interactions in an extra dimension,''
  JHEP {\bf 1410}, 191 (2014)
  [arXiv:1405.5872 [hep-ph]].

\bibitem{Fujimoto:2011kf}
  Y.~Fujimoto, T.~Nagasawa, S.~Ohya and M.~Sakamoto,
  ``Phase Structure of Gauge Theories on an Interval,''
  Prog.\ Theor.\ Phys.\  {\bf 126} (2011) 841
  [arXiv:1108.1976 [hep-th]].

\bibitem{Fujimoto:2013ki}
  Y.~Fujimoto, K.~Nishiwaki and M.~Sakamoto,
  ``CP phase from twisted Higgs vacuum expectation value in extra dimension,''
  Phys.\ Rev.\ D {\bf 88} (2013) 11,  115007
  [arXiv:1301.7253 [hep-ph]].

\bibitem{Witten:1981nf}
  E.~Witten,
  ``Dynamical Breaking of Supersymmetry,''
  Nucl.\ Phys.\ B {\bf 188} (1981) 513.

\bibitem{Rizzo:2006nw}
  T.~G.~Rizzo,
  ``$Z^\prime$ phenomenology and the LHC,''
  hep-ph/0610104.

\bibitem{Cheon:2000tq}
  T.~Cheon, T.~Fulop and I.~Tsutsui,
  ``Symmetry, duality and anholonomy of point interactions in one-dimension,''
  Annals Phys.\  {\bf 294} (2001) 1
  [quant-ph/0008123].

\bibitem{Nagasawa:2008an}
  T.~Nagasawa, S.~Ohya, K.~Sakamoto, M.~Sakamoto and K.~Sekiya,
  ``Hierarchy of QM SUSYs on a Bounded Domain,''
  J.\ Phys.\ A {\bf 42} (2009) 265203
  [arXiv:0812.4659 [hep-th]].

\bibitem{Cooper:1994eh}
  F.~Cooper, A.~Khare and U.~Sukhatme,
  ``Supersymmetry and quantum mechanics,''
  Phys.\ Rept.\  {\bf 251}, 267 (1995)
  [hep-th/9405029].

  F.~Cooper, A.~Khare and U.~Sukhatme,
  ``Supersymmetry in quantum mechanics,''
  Singapore, Singapore: World Scientific (2001) 210 p













\bibitem{GonzalezGarcia:2012sz}
  M.~C.~Gonzalez-Garcia, M.~Maltoni, J.~Salvado and T.~Schwetz,
  ``Global fit to three neutrino mixing: critical look at present precision,''
  JHEP {\bf 1212} (2012) 123
  [arXiv:1209.3023 [hep-ph]].

\bibitem{Xing:2011ur}
  Z.~z.~Xing,
  ``A full parametrization of the 6 X 6 flavor mixing matrix in the presence of three light or heavy sterile neutrinos,''
  Phys.\ Rev.\ D {\bf 85} (2012) 013008
  [arXiv:1110.0083 [hep-ph]].

\bibitem{Beringer:1900zz}
  J.~Beringer {\it et al.}  [Particle Data Group Collaboration],
  ``Review of Particle Physics (RPP),''
  Phys.\ Rev.\ D {\bf 86} (2012) 010001.

\bibitem{Dreiner:2008tw}
  H.~K.~Dreiner, H.~E.~Haber and S.~P.~Martin,
  ``Two-component spinor techniques and Feynman rules for quantum field theory and supersymmetry,''
  Phys.\ Rept.\  {\bf 494} (2010) 1
  [arXiv:0812.1594 [hep-ph]].

\bibitem{Gando:2012zm}
  A.~Gando {\it et al.}  [KamLAND-Zen Collaboration],
  ``Limit on Neutrinoless $\beta\beta$ Decay of Xe-136 from the First Phase of KamLAND-Zen and Comparison with the Positive Claim in Ge-76,''
  Phys.\ Rev.\ Lett.\  {\bf 110} (2013) 6,  062502
  [arXiv:1211.3863 [hep-ex]].

\bibitem{Appelquist:2002mw}
  T.~Appelquist, B.~A.~Dobrescu and A.~R.~Hopper,
  ``Nonexotic neutral gauge bosons,''
  Phys.\ Rev.\ D {\bf 68} (2003) 035012
  [hep-ph/0212073].

\bibitem{Coriano:2014mpa}
  C.~Coriano, L.~Delle Rose and C.~Marzo,
  ``Vacuum Stability in U(1)-Prime Extensions of the Standard Model with TeV Scale Right Handed Neutrinos,''
  arXiv:1407.8539 [hep-ph].


\end{thebibliography}
\end{document}